\documentclass[aps,pra,reprint,groupedaddress]{revtex4-2}

\usepackage{xcolor}
\usepackage{amsmath,amsfonts,amssymb,amstext,graphicx}
\usepackage[colorlinks=true,  
citecolor=blue, 
linkcolor=blue, 
urlcolor=black, 
anchorcolor = blue]{hyperref}

\usepackage{cancel}
\usepackage{empheq}
\usepackage{float} 
\usepackage{subcaption}
\usepackage{comment}
\usepackage{blindtext}
\usepackage{cases}
\usepackage{ulem}

\usepackage[bottom]{footmisc}
\usepackage{lineno}

\usepackage{dcolumn}
\usepackage{bm}
\usepackage{orcidlink}

\def\k#1{\mathinner{|{#1}\rangle}}
\def\bra#1{\mathinner{\langle{#1}|}}
\def\ket#1{\mathinner{|{#1}\rangle}}

\newcommand{\bk}[2]{\langle #1|#2\rangle}
\newcommand{\kb}[2]{|#1\rangle   \langle #2|}

\newcommand{\expt}[3]{\langle #1|#2|#3\rangle}

\newcommand{\tr}[1]{{{\rm Tr~}{#1}}}

\newcommand{\av}[1]{\langle #1 \rangle}

\newcommand{\sch}{{Schr\"{o}dinger}}

\begin{document}
	
\title{Tensor-Network-Based Unraveling of Non-Markovian Dynamics\\ in Large Spin Chains via the Influence Martingale Approach}

\author{Sujay Mondal \orcidlink{0009-0005-2696-2405} }
\email{sujay.mondal.phy23@gm.rkmvu.ac.in}

\author{Siddhartha Dutta \orcidlink{0009-0006-5975-0350}}
\email{siddhartha.dutta.phy23@gm.rkmvu.ac.in}

\author{Abhijit Bandyopadhyay \orcidlink{0009-0009-4113-3356}}
\email{abhijit.phy@gm.rkmvu.ac.in}

\affiliation{Department of Physics,\\Ramakrishna Mission Vivekananda Educational and Research Institute\\ Belur Math, Howrah-711202, West-Bengal, India}

\begin{abstract} 
 Classical simulation of open quantum system dynamics remains challenging due to the exponential growth of the Hilbert space, the need to accurately capture dissipation and decoherence, and the added complexity of memory effects in the non-Markovian regime.
  We develop an efficient algorithm for simulating both Markovian and non-Markovian dynamics in large one-dimensional quantum systems. Extending the Tensor Jump Method, which combines TDVP-based tensor-network evolution with a Suzuki--Trotter decomposition of stochastic trajectories, our approach incorporates time-dependent decay rates—treating positive rates as time-inhomogeneous Markovian processes and negative rates via the Influence Martingale formalism to unravel time-local non-Markovian dynamics. 
  {We further introduce the concept of the `influence radius' to achieve a resource-efficient framework enabling} scalable simulations of open-system dynamics in the non-Markovian regime, as demonstrated for a one-dimensional transverse-field Ising chain comprising up to 100 spin qubits.
\end{abstract}

 \maketitle

\section{Introduction}
\label{sec:intro} 
A closed quantum system is an idealized construct that 
provides a tractable framework for studying intrinsic 
quantum dynamics in the absence of environmental 
interactions and serves as the theoretical basis for 
exploring unitary evolution, coherence, and entanglement 
in quantum theory.  
In practice, quantum systems are never perfectly 
isolated; unavoidable environmental couplings--such as 
thermal and vacuum fluctuations, stray fields, or other 
background noise--induce decoherence and dissipation, 
rendering all systems inherently open 
\cite{Haroche2006ExploringTQ,Breuer2007,sym14081752}. 
The investigation of open quantum system dynamics is fundamentally driven by the need to understand how environmental interactions influence quantum behavior, giving rise to stochastic fluctuations and decoherence. 
From a technological perspective,these insights are critical for  characterizing and mitigating noise in quantum devices \cite{Myatt2000DecoherenceOQ,Khalil:2019rnx, Gottesman,Shaibali2023}
facilitating the design of robust, scalable, and high-fidelity quantum technologies 
that function reliably under realistic conditions.\\

The dynamics of open quantum systems are generally described by master equations, whose structure is determined by the underlying model of system-environment coupling. In the Markovian regime, where environmental correlations decay on timescales much shorter than those of the system, the reduced dynamics can be expressed in the  Gorini-Kossakowski-Sudarshan-Lindblad (GKSL) form  \cite{Lindblad1976, Gorini1976, Breuer2007}, ensuring complete positivity and trace preservation of the dynamical map for the system's reduced density operator through a time-local generator.
Beyond this regime, system--environment correlations persist and memory effects become significant, leading to reduced dynamics that are referred to as non-Markovian. 
Such dynamical behavior can be captured either by time-convolutionless master equations with time-dependent coefficients that may temporarily take negative values, or by integro-differential equations incorporating explicit memory kernels \cite{Breuer2007, Breuer2016}.
Distinguishing between Markovian and non-Markovian regimes is therefore essential for a consistent characterization of decoherence and dissipation in open quantum systems, as well as for developing strategies to harness or mitigate these effects in quantum technologies. 
 Exact analytical solutions of master equations, however, are limited to highly idealized system--environment 
 interaction models. In realistic settings involving many-body systems, structured reservoirs, or pronounced non-Markovian effects, the resulting master equations become structurally complex and generally analytically intractable.
 This necessitates the use of classical computational approaches to simulate the reduced dynamics.  
 These methods enable a systematic study of decoherence, dissipation, and the evolution of quantum correlations, and provide a framework to benchmark emerging quantum technologies under realistic noise conditions \cite{weimersimulation2021}.\\

However, the simulation of open quantum systems on classical computers remains highly challenging due to the intricate structure inherent in their dynamics, requiring the development and application of specialized computational methods to address these challenges \cite{PhysRevLett.93.207204,PhysRevLett.93.207205,Werner2016,PhysRevA.92.052116,Rotureau:2006rf,Tamascelli2015,  McKeever2022}.
In particular, the exponential growth of the Hilbert 
space with system size imposes significant constraints 
on many-body simulations, which can be efficiently 
addressed using tensor-network methods such as matrix product states (MPS) \cite{PhysRevA.82.063605,transchel2014montecarlotimedependentvariational,Purkayastha2020,Purkayastha2021,Moroder2023,Xie2024,Jaschke2018,PRXQuantum.5.010308,NewStoudenmire}. These methods encode many-body states as networks of low-rank tensors while preserving the essential entanglement structure. 
Nevertheless, accurately tracking entanglement and 
quantum correlations in large-scale systems remains 
computationally challenging, as their growth during time 
evolution leads to rapidly escalating 
resource requirements \cite{SchuchQuench,AnalyticalQuench,Orus2019,Schollwock2011,PhysRevLett.114.220601,PhysRevLett.111.150403}.  
 To address the challenges of simulating large-scale quantum systems, efficient tensor-network algorithms such as Time-Evolving Block Decimation (TEBD) \cite{Vidal2004,Tamascelli2015}, Density Matrix Renormalization Group (DMRG) \cite{Daley2004,Schollwock2011,Rotureau:2006rf}, and the Time-Dependent Variational Principle (TDVP) \cite{Haegeman2011,PhysRevLett.100.130501,Haegeman2016,Paeckel2019, PhysRevLett.127.230501} have been developed. 
TEBD enables efficient time evolution of one-dimensional 
systems with short-range interactions through Trotter 
decomposition within the MPS framework. DMRG allows 
highly accurate determination of ground and low-lying 
excited states by variationally optimizing MPS and 
systematically truncating the Hilbert space using 
reduced density matrices. TDVP extends the MPS formalism 
by evolving states within a constrained variational 
manifold, dynamically adjusting the bond dimension to 
balance accuracy and computational cost, thereby 
facilitating accurate simulations of correlated quantum 
dynamics over extended timescales.
Simulating non-Markovian dynamics, which arise from memory effects and time-nonlocal correlations, typically requires tracking the system’s history, incurring substantial computational cost. Under weak-to-moderate system-bath coupling, factorized initial conditions, time-translation-invariant bath correlations, and finite bath correlation times, the non-Markovian master equation with explicit memory kernels can be recast in a time-convolutionless (TCL) form \cite{PhysRevA.97.012127,Breuer2016}. This time-local formulation incorporates memory effects into time-dependent rates, which may temporarily become negative, indicating a backflow of information \cite{Breuer2016, PhysRevLett.103.210401, PhysRevLett.105.050403}. By eliminating the need to retain the full system history, the TCL approach significantly reduces computational overhead \cite{PhysRevLett.104.070406}.\\

Considerable progress in modeling and simulating Markovian quantum dynamics has been enabled by specialized numerical toolkits, most notably the Quantum Toolbox in Python (QuTiP) \cite{johanssonqutip2012,johanssonqutip2013}. QuTiP 
integrates analytical master equation solvers with the stochastic Monte Carlo Wave-Function (MCWF) approach \cite{Molmer92,PhysRevA.46.4382,Dalibard1992,Carmichael1993,DaleyReview, PhysRevLett.133.230403,PhysRevLett.128.243601,Vovk2024}, enabling open-system dynamics to be simulated through ensembles of pure-state trajectories whose averages reproduce mixed-state evolution.
Although statistical convergence requires large trajectory
ensembles, computational costs can be reduced through importance sampling, adaptive time-stepping, and parallelization strategies \cite{KORNYIK201988,SciPostPhys.11.3.048,carino2003}.
These advances have facilitated detailed studies of dissipation, decoherence, and quantum control across a wide range of platforms, establishing QuTiP as an indispensable tool for both research and pedagogy. However, its reliance on explicit matrix representations of operators and density matrices leads to exponential scaling with system size, thereby limiting its applicability to small systems of few qubits.\\

 The Tensor Jump Method (TJM), introduced in the Munich Quantum Toolkit \cite{sander2025largescalestochasticsimulationopen}, is an efficient emerging framework for large-scale Markovian simulations of one-dimensional chains with limited entanglement growth.
This method combines TDVP-based MPS unitary evolution with a Suzuki-Trotter decomposition of stochastic Monte Carlo trajectories. By incorporating dissipative processes and quantum jumps with strictly positive decay rates, TJM provides a scalable unraveling of the Markovian GKSL master equation.
In contrast, trajectory-based approaches for non-Markovian dynamics are comparatively less developed due to the negative-probability issue \cite{PhysRevA.79.062112}, although recent studies have increasingly explored methods for non-Markovian simulations \cite{oqupy,PhysRevResearch.5.033078,acs.jpclett.4c03431}.
 One of the most promising approaches involves the use of the influence martingale, formulated within the quantum-trajectory framework for general time-local master equations \cite{donvil2022}. 
This method implements a time-local master equation with non-Markovian unraveling and has been demonstrated using QuTiP's Monte Carlo solver to evolve the quantum trajectories of qubit chains containing up to 13 qubits \cite{donvil2022}.\\

 Simulating quantum dynamics in (moderately) 
large-scale systems, such as spin chains consisting 
of on the order of 50-100 qubits, is essential for probing collective many-body phenomena that remain inaccessible in smaller setups. 
While short chains provide valuable benchmarks for controlled studies of open-system dynamics, only sufficiently long systems exhibit phenomena such as thermalization, entanglement spreading and the interplay between quantum correlations and environmental couplings that underlie non-Markovian relaxation and dissipative transitions.
As experimental platforms continue to scale,
efficient simulation methods capable of capturing both many-body correlations and non-Markovian effects are essential for understanding complex quantum dynamics and guiding the development of scalable quantum technologies. \\

In this work, we present an algorithm designed for the efficient 
simulation of both Markovian and non-Markovian open-system dynamics 
in large-scale one-dimensional quantum systems, and demonstrate its 
application to spin chains of up to 100 qubits.
We employ the Tensor Jump Method  
\cite{sander2025largescalestochasticsimulationopen}, 
which embeds stochastic quantum jumps into MPS tensor networks to 
efficiently unravel GKSL master equations, extended to account for 
time-dependent decay rates that may become temporarily negative 
over certain intervals.
When decay rates remain strictly positive, 
the method enables the simulation 
of realistic, time-inhomogeneous Markovian dynamics. For decay 
rates that become temporarily negative, we utilize the influence 
martingale formalism \cite{donvil2022}, allowing the unraveling of 
non-Markovian master equations expressed in the 
time-local form. {we introduce the concept of `influence radius', which serve as a quantifier of the local extent to which martingale corrections for local observables are needed to be implemented.}
Combined with TDVP-based tensor-network evolution and stochastic 
quantum jumps within the TJM, this approach renders simulations of 
non-Markovian dynamics in large-scale systems computationally 
feasible, while maintaining efficient control over computational 
resources. This unified framework thus supports scalable 
simulations of open-system dynamics, encompassing both time--inhomogeneous Markovian processes and genuinely non-Markovian 
behavior, as demonstrated for a one-dimensional transverse-field 
Ising chain with system sizes of up to 100 spin qubits.\\ \\
The remainder of this article is organized as follows. In 
Sec.~\ref{sec:master}, we provide a concise overview of master 
equations governing both Markovian and non-Markovian dynamics. 
Building on this foundation, Sec.~\ref{sec:unravelmark} discusses 
stochastic unravelings in the Markovian regime, which are then 
extended to the non-Markovian scenario in 
Sec.~\ref{sec:unravelnonmark}. In Sec.~\ref{sec:network}, we 
outline how these unraveling schemes can be implemented within the TJM framework. 
Sec.~\ref{sec:tjm} presents a detailed description of the TJM, 
covering the initialization scheme, Trotterized evolution including 
both coherent and dissipative processes (Sec.~\ref{sec:tjm-trot}), 
the treatment of stochastic jumps (Sec.~\ref{sec:tjm-jump}), {
the implementation of unitary dynamics via the dynamic TDVP within 
tensor networks (Sec.~\ref{sec:tjm-tdvp}) and discussion on Computational complexity of this approach (Sec.~\ref{sec:ccomplexity}).}  
In Sec.~\ref{sec:nu}, we present and discuss numerical results {on a representative one-dimensional spin chain of large system size: benchmarks in comparison with MPO-based numerically exact method (Sec.~\ref{sec:bmrkn}), and introduction and application of `Influence radius' (Sec.~\ref{sec:infrad}), that establish the regime of relevance for exploratory research and demonstrate the scalability of the proposed algorithm.} The conclusions are summarized in Sec.~\ref{sec:con}.

\section{Master Equations for Markovian and Non-Markovian Dynamics}
\label{sec:master} 
Unlike isolated systems, which evolve unitarily under their Hamiltonian in Hilbert space, open quantum systems are described by a reduced density matrix formalism that accounts for decoherence and entanglement arising from interactions with the environment.
The evolution of  reduced density operator
of an open quantum system is generally described by a quantum master equation, comprising a unitary term that generates coherent dynamics under the system Hamiltonian and a non-unitary term that captures dissipation and decoherence arising from interactions with the environment.
Depending on the presence of memory effects--the influence of a 
system's past states on its future evolution--open-system dynamics 
are broadly classified as Markovian, where information flows 
irreversibly to the environment and the dynamics are effectively 
memoryless, or non-Markovian, where information backflow induces 
history-dependent evolution, typically arising from strong 
coupling, structured environments, or initial system-environment 
correlations.\\

\textbf{Time-homogeneous Markovian dynamics:} 
Time-homogeneous Markovian dynamics corresponds to the simplest scenario of memoryless open-system evolution, where the environment is stationary and the rates governing dissipation and decoherence remain constant in time.  
In this setting, the reduced density operator $\rho$ of the system evolves according to a time-homogeneous master equation of the  GKSL  form  \cite{Gorini1976,Lindblad1976}, commonly known as the Lindblad equation, with a time-independent superoperator $\mathcal{L}$ (Lindbladian) generating the dynamics: 
\begin{eqnarray} 
	\label{eq:ss1}
 \frac{d\rho(t)}{dt} &=&   \mathcal{L}[\rho(t)] = -i[H_0, \rho(t)] \\ \nonumber
  &&+ \sum_k \gamma_k \left( L_k \rho(t) L_k^\dagger - \frac{1}{2} \left\{ L_k^\dagger L_k, \rho(t) \right\} \right)\,,
\end{eqnarray}
where  
we set $\hbar = 1$ and the Hamiltonian $H_0 \equiv H_S + H_{LS}$
combines the bare system Hamiltonian $H_S$
with a constant Lamb-shift term $H_{LS}$
arising from virtual excitations exchanged with the environment \cite{Breuer2007}.
This coherent, non-dissipative renormalization of the system Hamiltonian, arising from the 
imaginary part of the bath correlation functions, modifies the phase evolution of the states 
without inducing irreversible energy transfer. In the Markovian limit, the constant Lamb-shift term captures stationary bath-induced level shifts, 
reflecting a time-independent coherent contribution to the dynamics.
The parameter $\gamma_k$ denotes the constant, positive decay rate of the
$k$-th environmental channel,  while $L_k$ is the corresponding Lindblad (jump) operator encoding  the environmental effects on the system. 
The derivation of Eq.~\eqref{eq:ss1} typically stems from a microscopic system-environment interaction model, under the following assumptions:  
(i) The initial state is factorized between system and environment;  
(ii) Weak coupling (Born approximation) -  allowing perturbative treatment up to second order;  
(iii) Markov approximation, assuming fast decay of environmental correlations compared to the system's evolution timescale;  
(iv) Rotating Wave Approximation -  where rapidly oscillating terms average out and are neglected;  
(v) Environmental stationarity - typically modeled as a time-invariant  state, leading to time-translation invariant correlation functions,  
$\langle B_\alpha(t) B_\beta(t') \rangle = \langle B_\alpha(t - t') B_\beta(0) \rangle$, ($B_{\alpha,\beta}$ are the
operators on the Hilbert space of the environment)  
implying time-independent decay rates $\gamma_k$.  \\

The Lindblad master Eq.~\eqref{eq:ss1}   generates Completely Positive Trace-Preserving (CPTP) dynamical maps   $\Phi(t,0)$  \cite{PhysRevLett.105.050403}, that evolve the system state from the initial time $t=0$ to any later time $t$. 
 In the Markovian regime, reflecting the absence of memory effects,  
$\Phi$ is   CP-divisible satisfying 
\begin{eqnarray}
{\Phi(t,0) = \Phi(t,s)\Phi(s,0), \quad \text{for all } 0 \leq s \leq t,}
\label{eq:ss2}
\end{eqnarray}
so that the evolution can be represented as a consistent sequence of intermediate CPTP transformations, preserving the structure of a quantum Markov process.\\

{
In realistic open-system settings, however, the effective generator may acquire
an explicit time dependence due to structured reservoirs, finite bath
correlation times, external driving, or time-dependent system--environment
couplings. A concrete example is provided by a driven two-state system coupled
to a generic structured reservoir, for which the TCL master equation
contains explicitly time-dependent decay coefficients determined by the
reservoir spectral density~\cite{PhysRevA.81.052103}. Similar time-dependent rates arise in optical and cavity-QED settings. For instance, in
Ref.~\cite{Salatinoeurphys}, dissipative photonic dynamics is described by an
amplitude-damping channel in which the time-dependent decay rate is obtained
from the excited-state probability amplitude. For an atom-cavity system with cavity losses, non-flat spectral profiles lead to non-exponential, time-dependent and
non-Markovian decay dynamics~\cite{scala2008}. Furthermore, the spontaneous
decay of an atom coupled to a reservoir can be controlled by manipulating the
reservoir mode frequencies, leading to controllable, explicitly time-dependent effective decay dynamics \cite{Linington2006}. Thus it is intriguing to explore both the time-inhomogeneous Markovian and non-Markovian dynamics.}\\

\textbf{Time-inhomogeneous Markovian dynamics:} 
The Markovian character of open quantum dynamics is fundamentally determined by the 
CP-divisibility of the dynamical map, rather than by the time-independence of the 
Lindbladian $\mathcal{L}$. 
When environmental influences or external control protocols induce explicit 
time dependence in the generator $\mathcal{L}$, the decay rates 
 may vary in time, while remaining non-negative for all time $t$. As long as this condition is satisfied, the dynamics remain  CP-divisible and thus retain their Markovian character:
  information flows irreversibly from the system to the environment without backflow, even if the   rates $\gamma_k(t) > 0$ vary in time, giving rise to time-inhomogeneous Markovian dynamics. \\

\textbf{Non Markovian dynamics:} 
In many realistic scenarios--such as structured or finite reservoirs, spin environments, low temperatures, or strong system-environment coupling--system-environment interactions give rise to memory effects. The environment retains information about the system's past and can feed it back  during evolution, leading to non-Markovian dynamics, where the standard GKSL master equation no longer provides an adequate dynamical description. Two key frameworks for deriving master equations in such regimes are the Nakajima-Zwanzig (NZ) and  TCL  formalisms \cite{Breuer2007}. Both employ projection operator techniques but differ fundamentally in how memory and time dependence are incorporated
into the reduced system dynamics.\\

The NZ approach   leads to a   integro-differential equation of the  reduced density matrix in the following form \cite{10.1143/PTP.20.948,sym14081752}:
\begin{eqnarray}
{\frac{d}{dt} \rho(t) = \int_0^t K(t - \tau) \, \rho(\tau) \, d\tau,}
\label{eq:ss3}
\end{eqnarray}
where $K(t - \tau)$  
 is the memory kernel that incorporates the entire history of the system. The resulting Eq.\ \eqref{eq:ss3} is non-local in time, making it well-suited to describe strong memory effects.  However, the presence of the time convolution complicates both the analytical treatment and numerical implementation.
In contrast, the TCL approach 
yields a time-local master equation
\begin{eqnarray}
\frac{d}{dt} \rho(t) = {\cal L}_t[ \rho(t)]
\label{eq:ss4}
\end{eqnarray}
for the reduced density matrix, where the time-dependent 
superoperator ${\cal L}_t$ captures the 
cumulative influence of the environment up to  
current time $t$, without depending explicitly on the system's 
state $\rho(\tau)$ at earlier times   $\tau < t$.
non-Markovianity is retained through the explicit 
time dependence of ${\cal L}_t$, allowing the treatment 
of information backflow and structured reservoirs in a more tractable form.
The TCL approach involves constructing a perturbative
expansion  in the system-environment coupling, typically truncated at second order. 
Validity of this approach rests on premises of weak-to-moderate coupling, factorized initial conditions,
time-translation-invariant correlation functions, and finite bath correlation times to ensure convergence.
Under these approximations, the TCL master equation assumes a structure   formally analogous to the Lindblad equation \cite{Lindblad1976}:

\begin{eqnarray} 
\label{eq:ss5}
\frac{d}{dt} \rho(t) &=& {\cal L}_t[ \rho(t)]
= -i[H_0(t), \rho(t)] \\ \nonumber
&& + \sum_k \gamma_k(t) \left( L_k \rho(t) L_k^\dagger 
- \frac{1}{2} \left\{ L_k^\dagger L_k, \rho(t) \right\} \right)\,,
\end{eqnarray} 

where $H_0(t) \equiv H_S + H_{LS}(t)$ is the system Hamiltonian incorporating the time dependent Lamb shift $H_{LS}(t)$, arising from the finite memory of the environment.
In this framework, the time dependence of $H_{LS}(t)$ accounts for dynamical modulation of coherent oscillations and transient energy-level shifts, 
representing environment-induced coherent back-action that complements the dissipative non-Markovian effects encoded in the decay rates $\gamma_k(t)$ in Eq.~\eqref{eq:ss5}.
In contrast to the Markovian case, these time-dependent decay rates may 
temporarily attain negative values over finite intervals. Such sign 
reversals correspond to partial restoration of populations and coherences, 
indicating information backflow and a breakdown of CP-divisibility.  \\

Owing to the boundedness of the time-local generator $\mathcal{L}_t$, 
constructed perturbatively from finite and smooth bath correlation functions, 
the propagator
$\Phi(t,0) = \mathcal{T} \exp(\int_0^t \mathcal{L}_\tau\, d\tau)$ 
($\mathcal{T}$ denotes the time-ordering operator), 
remains bounded. 
In the non-Markovian regime, however, negative decay rates $\gamma_k(t)$ 
induced by memory effects break CP-divisibility
implying that intermediate propagators $\Phi(t+\tau,t)$ 
may not be completely positive, even though $\Phi(t,0)$ 
still produces valid states for the relevant initial conditions. 
Consequently, in the TCL framework, the dynamical map $\Phi(t,0)$ is linear, 
trace-preserving, and completely bounded for all finite times 
due to the regularity of its generator, but loses CP-divisibility as a result of information backflow \cite{Breuer2016}.

\section{Stochastic Unraveling  for Markovian Evolution}
\label{sec:unravelmark}
Simulation of an open quantum system refers to the numerical modeling of the reduced system dynamics 
 in the presence of environmental interactions,
typically through relevant master equations or equivalent stochastic formulations without explicitly resolving the multitude of environmental modes.
The objective is to faithfully capture the essential features of the dynamics while ensuring computational feasibility. 
Such simulations are often computationally demanding, 
 which has motivated the development of a variety of numerical techniques--each with specific advantages and limitations--for solving the master equations governing open system evolution \cite{CiracZollerSimulation,PhysRevResearch.3.023005,weimersimulation2021}. \\
 
In the Markovian regime, a direct approach to simulating the GKSL 
master Eq.~\eqref{eq:ss1} consists of explicitly constructing 
the Lindbladian superoperator $\mathcal{L}$ and numerically 
integrating the resulting equation for the density matrix $\rho$. 
To facilitate this, $\rho$ is  vectorized  into a column 
vector of length $d^2$ for a Hilbert space of dimension $d$, 
allowing the GKSL equation to be recast as a standard matrix-vector 
differential equation governed by a $d^2 \times d^2$ Lindbladian 
matrix. This approach is conceptually simple and yields numerically 
exact results within machine precision, providing a direct and transparent 
method for simulating open-system dynamics. However, its 
computational complexity scales unfavorably with system size, as 
both memory usage and runtime increase rapidly with $d^2$, limiting 
its applicability to relatively small Hilbert spaces despite its 
accuracy.\\

An alternative to direct density matrix evolution is to \textit{unravel} the master equation into an ensemble of stochastic pure-state trajectories, whose statistical average reproduces the density matrix dynamics.
This approach reformulates the open-system evolution as a stochastic Schr\"odinger equation, implemented via discrete quantum jumps in the Monte Carlo Wave-function 
(MCWF) method \cite{Molmer92,PhysRevA.46.4382,Dalibard1992}.
The unraveling lowers computational cost by evolving $d$-dimensional state vectors instead of $d^2$-dimensional density matrices and provides a physical interpretation of individual trajectories as possible realizations of the system's evolution.
 However, reconstructing the density matrix accurately requires averaging over many trajectories, which can
increase the overall computational effort.\\

For Markovian dynamics described by the GKSL master equation (with $\gamma_k>0$ for all $t$), MCWF method evolves the system in discrete time steps $\delta t$, propagating a pure state $\lvert \psi(t) \rangle$ from time $t$ to $t + \delta t$.
At each time step, the state  
either evolves deterministically under the non-unitary operator $e^{-i H_{\rm eff} \delta t}$, generated by the effective non-Hermitian Hamiltonian $H_{\rm eff} = H_0 + H_D$ where $H_D \equiv - \frac{i}{2} \sum_k \gamma_k L_k^\dagger L_k$ 
or,  experiences a stochastic quantum jump.  
For sufficiently small time steps $\delta t$, the deterministic
 evolution propagates the state $\k{\psi(t)}$ to 
\begin{eqnarray}
{\k{\psi(t+\delta t)} }
&=& 
 {\left( 1 - \frac{iH_{\rm eff}\delta t}{\hbar} \right) \k{\psi(t)}.}
\label{eq:ss7}
\end{eqnarray}
The term $H_D$ in the effective Hamiltonian $H_{\rm eff}$, which involves the jump operators $L_k$, reduces the norm of $\lvert \psi(t) \rangle$, reflecting population loss due to dissipation through quantum jumps into various environmental channels.
For a normalized state  $\k{\psi(t)}$, the deviation from unit norm after a small time step $\delta t$
can be quantified by the denormalization factor 
\begin{eqnarray}
\delta p(t) = 1 - \bk{\psi(t+\delta t)}{\psi(t+\delta t)},
\label{eq:ss8}
\end{eqnarray}
which provides a stochastic criterion for determining whether a quantum jump occurs within the interval $\delta t$.
The probability that the state $|\psi(t)\rangle$ experiences no stochastic quantum jump during the interval $[t, t+\delta t]$, and thus evolves deterministically, is given by $1 - \delta p(t)$.
If a quantum jump into channel $k$ occurs during the time step $\delta t$, the state $\k{\psi(t)}$
is instantaneously transformed to $L_k\k{\psi(t)}$
with  a reduced norm $||L_k\k{\psi(t)}|| \equiv \sqrt{\expt{\psi(t)}{L_k^\dagger L_k}{\psi(t)}}$,
whose square  corresponds to the occupation probability of the decaying state in  channel $k$. 
The probability $\delta p_k$, of a quantum jump to channel $k$ within the interval $[t, t+\delta t]$ is proportional to 
 this occupation probability
and also to the time step size ($\delta t$), with the proportionality constant being the decay rate $\gamma_k$: $
\delta p_k = \gamma_k\,\delta t\,\expt{\psi(t)}{L_k^\dagger L_k}{\psi(t)}$.
The denormalization factor $\delta p(t)$  is the sum of the individual stochastic contributions from all jump processes: 
$\delta p(t) = \sum_k \delta p_k(t)$. 
The deterministic evolution, occurring with probability $1 - \delta p(t)$, together with stochastic quantum jumps in the various channels $k$, each with probability $\delta p_k(t)$, transforms the pure state $\lvert \psi(t) \rangle$ at time $t$ into an ensemble of states at $t + \delta t$. For sufficiently small time steps, the infinitesimal change in the density operator, obtained by averaging over many such trajectories, reproduces the evolution described by the GKSL master equation \cite{Molmer92,PhysRevA.46.4382,Dalibard1992}. This correspondence establishes that the  MCWF  method provides an equivalent description of Markovian open-system dynamics while offering a computationally efficient and physically intuitive framework.\\

The simulation algorithm proceeds as follows. 
A random number $\epsilon \in [0,1]$, drawn from a uniform 
distribution, is generated and compared with $\delta p$. If $
\epsilon \geq \delta p$, no quantum jump occurs, and the 
 state $\ket{\psi(t+\delta t)}$ 
deterministically evolves to 
state $\ket{\psi(t+\delta t)} = 
e^{-iH_{\rm eff}\delta t} \k{\psi(t)}$ before advancing to the next time step. 
If $\epsilon < \delta p$, a stochastic quantum jump is triggered. 
The jump channel $k$ is selected from the set of
all channels $\{k\}$ with 
probability $\delta p_k / \delta p$, ($\sum_k \delta p_k 
/ \delta p = 1$). The corresponding jump operator $L_k$ is then 
applied directly to the state $\ket{\psi(t)}$, followed by 
normalization, yielding the post-jump state
\begin{eqnarray}
\frac{L_k \ket{\psi(t)}}{\sqrt{\bra{\psi(t)} L_k^\dagger L_k \ket{\psi(t)}}}
= \frac{\sqrt{\gamma_k \delta t}\, L_k \ket{\psi(t)}}{\sqrt{\delta p_k}}.
\label{eq:ss12}
\end{eqnarray}
During the deterministic (no-jump) evolution under the 
non-Hermitian Hamiltonian $H_{\rm eff}$, 
the state is intentionally left unnormalized, 
as the decaying norm encodes the probability 
that the trajectory continues without undergoing a quantum jump. 
In contrast, when a jump occurs, 
the state is abruptly transformed by a non-norm-preserving 
jump operator. Since the jump probability is already accounted for in the stochastic sampling, the post-jump state is subsequently 
renormalized to produce a valid quantum state for further evolution. 
This procedure is iterated until the desired total evolution time 
$T$ is reached, yielding a single quantum trajectory and a final 
state vector $\ket{\psi(T)}$. Within the MCWF framework, the open-
system dynamics are thus unraveled into stochastic trajectories of 
the pure state $\ket{\psi(t)}$, each consisting of deterministic 
non-unitary evolution  interspersed with  random quantum jumps. 
Ensemble averaging over a sufficiently large number of such 
trajectories reproduces the Markovian dynamics of the system as 
governed by the GKSL master equation, without requiring direct 
integration of the density matrix evolution.\\

A mathematical formulation equivalent to the 
jump-type unraveling employed 
in the  MCWF  method can be expressed using the 
It\^o stochastic Schr\"odinger equation   (SSE)
\cite{Hudson1984} with discrete jumps:
\begin{eqnarray}
\label{eq:itom}
d|\psi(t)\rangle &=& - i H_0 |\psi(t)\rangle\, dt \nonumber \\ 
&& - \sum_k \gamma_k(t) \, \frac{L_k^\dagger L_k - \| L_k |\psi(t)\rangle \|^2}{2} \, |\psi(t)\rangle \, dt \nonumber \\ 
&& + \sum_k dN_t^{(k)} \left( \frac{L_k |\psi(t)\rangle}{\| L_k |\psi(t)\rangle \|} - |\psi(t)\rangle \right),
\end{eqnarray}
where $\gamma_k(t) \ge 0$ for all $t$, ensuring Markovian dynamics. 
 $dN_t^{(k)}$ denotes the Poisson increment associated with the 
jump channel $k$ at instant $t$. The increment takes values of either 0 or 1, serving as a counter 
for jumps along a single trajectory, and satisfies  
$dN_t^{(k)} dN_t^{(k')} = \delta_{kk'} dN_t^{(k)}$.
The expectation
$\mathbb{E}\big[ dN_t^{(k)} \big] = \gamma_k(t)\, \| L_k |\psi(t)\rangle \|^2 dt$,  represents the probability of a quantum jump occurring in channel $k$ within the time interval $dt$.
The It\^o SSE  \eqref{eq:itom}  captures the discrete quantum jumps observed in MCWF  simulations, while the ensemble average over many trajectories reproduces 
the corresponding GKSL equation. \\

\section{Stochastic Unraveling  for Non-Markovian Evolution}
\label{sec:unravelnonmark}
 The standard It\^o SSE in Eq.\ \eqref{eq:itom} is derived under the condition that all decay rates satisfy $\gamma_k(t) \geq 0$ for all times, which guarantees CP-divisibility and corresponds to Markovian dynamics. When the rates $\gamma_k(t)$ become temporarily negative, the dynamics enter the non-Markovian regime \cite{PhysRevA.79.062112}. In such cases, standard unraveling
 of the non-Markovian master equation in GKSL-analogous form as in Eq.\ \eqref{eq:ss5} fails, since the expectation
$\mathbb{E}\big[ dN_t^{(k)} \big] = \gamma_k(t)\, \| L_k |\psi(t)\rangle \|^2 \, dt$, can no longer be consistently interpreted as a jump probability  in the presence of negative $\gamma_k(t)$. \\

To address this challenge,  one may invoke  the probabilistic framework of martingales (see \cite{donvil2022} and references therein), which provides a rigorous foundation for redefining stochastic processes in scenarios where the conventional Poisson interpretation breaks down. 
In probability theory, a martingale is a stochastic process whose conditional expectation at any future time, given the past history, coincides with its present value. This property implies that, conditioned on the current state, the process exhibits no systematic drift in future expectation, thereby serving as a canonical baseline in stochastic analysis against which modifications of probability measures can be consistently defined. Girsanov's theorem \cite{girsanov}, a cornerstone of stochastic analysis, enables the reformulation of otherwise intractable processes by shifting their complexity from the stochastic dynamics to a martingale weight that modifies the underlying probability measure. In the context of quantum trajectories, this role is fulfilled by the influence martingale, which reweights the reference probability measure associated with the counting processes. Through this construction, the complexity of non-Markovian modifications is absorbed into the weighting factor, ensuring that ensemble averages over the reweighted trajectories reproduce the correct physical dynamics  \cite{donvil2022}.
Specifically, for any observable or trajectory-dependent quantity $X_t$, the physical expectation under the true non-Markovian dynamics  governed by Eq.\ \eqref{eq:ss5} can be expressed as an ensemble average over the reference trajectories weighted by the influence martingale $\mu_t$:
\begin{eqnarray}
\mathbb{E}_{\rm phys}[X_t] = \mathbb{E}_{\rm ref}[\mu_t \, X_t]  \quad \mbox{with} \quad \mathbb{E}_{\rm ref}[\mu_t] = 1\,,
\label{eq:mart1}
\end{eqnarray}
where $\mathbb{E}_{\rm phys}$ and $\mathbb{E}_{\rm ref}$
respectively denote averaging under the physical (non-Markovian) measure and the reference (easier-to-simulate) Poisson measure.
The influence martingale ensures both trace preservation and faithful reproduction of the underlying non-Markovian   dynamics governed by   Eq.\ \eqref{eq:ss5} in TCL framework,
provided that the set of Lindblad operators $ L_{k}$ 
satisfy {the normalization condition : $\sum_{l} {L_l}^\dagger L_l = \mathbb{I}$, where $l$ may represent single or a group of Lindblad operators corresponding to the same decay rate. For this particular model, we choose the groups for individual sites}.
This idea can be employed to address the complications arising 
from temporarily negative decay rates $\gamma_k(t)$, 
which invalidate the  
standard unraveling using  Eq.~\eqref{eq:itom}. Rather than directly 
computing the trajectory evolution as a stochastic average, we 
introduce a shifted set of strictly positive rates, 
$r_k(t) = \gamma_k(t) + C_t$, 
where the universal shift $C_t$ is applied uniformly to all channels $k$ at a given $t$. 
The corresponding stochastic trajectory dynamics is then given by  
\begin{eqnarray}
\label{eq:itom2}
d|\psi(t)\rangle &=& - i H_0 |\psi(t)\rangle\, dt \nonumber \\
&& - \sum_k r_k(t) \, \frac{L_k^\dagger L_k - \| L_k |\psi(t)\rangle \|^2}{2} \, |\psi(t)\rangle \, dt \nonumber\\
&& + \sum_k dN_t^{(k)} \left( \frac{L_k |\psi(t)\rangle}{\| L_k |\psi(t)\rangle \|} - |\psi(t)\rangle \right),
\end{eqnarray}  
with $r_k(t) > 0$ ensuring that the conditional expectation  
\begin{eqnarray}
\mathbb{E}\!\left[ dN_t^{(k)} \,\big|\, |\psi(t)\rangle \right] 
= r_k(t)\, \| L_k |\psi(t)\rangle \|^2 \, dt
\label{eq:itom3}
\end{eqnarray}
consistently represents the probability of a quantum jump 
occurring in channel $k$ within  interval $dt$.
To compensate for the artificial modification of jump rates, 
which is introduced to construct a reference ensemble with a 
consistent probabilistic interpretation, the influence martingale
$\mu_t$ is  employed in accordance with Eq.~\eqref{eq:mart1},
and defined   through 
\begin{eqnarray}
{\rho(t) = \mathbb{E}_{\rm phys}\!\left[\,|\psi(t)\rangle\langle\psi(t)|\,\right] 
= \mathbb{E}_{\rm ref}\!\left[\,\mu_t \, |\psi(t)\rangle\langle\psi(t)|\,\right].}
\label{eq:mart2}
\end{eqnarray}
In this way, $\mu_t$ establishes the equivalence between the physical and reference ensemble descriptions by connecting the corresponding probability measures, thereby ensuring trace preservation and furnishing a consistent framework
for  non-Markovian unraveling. 
 The influence martingale satisfies $\mathbb{E}_{\rm ref}\big{[}\mu_t   \big{]} = 1$  for all $t$ with $\mu_0 = 1$  and 
evolves randomly in time  depending on the underlying reference stochastic trajectory
and its evolution is typically described by the stochastic differential equation  \cite{donvil2022}
{
\begin{eqnarray}
\label{eq:mart3}
d\mu_t &=& \mu_t \sum_k \left(\frac{\gamma_k(t)}{r_k(t)} - 1\right) \nonumber \\  && \Big{(}dN^{(k)}_t - r_k(t) || L_k \k{\psi(t)}||^2 dt\Big{)}.
\end{eqnarray}
}
It follows that   $\rho(t) = \mathbb{E}_{\rm ref}\Big{[}\mu_t \kb{\psi(t)}{\psi(t)} \Big{]}$ satisfies  the TCL master equation 
for non-Markovian dynamics as given in Eq.\ \eqref{eq:ss5}. 
In this way, the non-Markovian dynamics are reconstructed by evolving reference trajectories with shifted rates, while the influence martingale $\mu_t$ provides the appropriate reweighting to ensure consistency with the physical ensemble.   \\

The increment of the martingale, $d\mu_t$,
 can be decomposed into a continuous and a discrete contribution:
  $d\mu_t = {d\mu_t}^{\rm cont} + {d\mu_t}^{\rm dis}$, where
 the continuous part 
(${d\mu_t}^{\rm cont}$) and  the
discrete part (${d\mu_t}^{\rm dis}$) are respectively given by 
{
\begin{eqnarray}
{d\mu_t}^{\rm cont} 
&=&
- \mu_t \sum_k \left(\frac{\gamma_k(t)}{r_k(t)} - 1\right) \nonumber \\ 
&& \quad \cdot \quad  r_k(t) || L_k \k{\psi(t)}||^2 dt,
\label{eq:mart5a} \\
{d\mu_t}^{\rm dis} 
&=&
 \mu_t \sum_k \left(\frac{\gamma_k(t)}{r_k(t)} - 1\right)  dN^{(k)}_t.
\label{eq:mart5b} 
\end{eqnarray}
}
The continuous component 
accounts for drift corrections arising from the 
discrepancy between the original decay rates 
$\gamma_k(t)$ and 
the shifted positive rates $r_k(t)$.
This term governs the deterministic evolution of $\mu_t$ and encodes a continuous reweighting of the reference trajectories.
The discrete component arises from the jump contributions 
$dN^{(k)}_t$ and compensates for the statistical discrepancy between jumps generated with the shifted rates $r_k(t)$ and those governed by the physical rates $\gamma_k(t)$.
 Together, these contributions ensure that the martingale-weighted trajectory ensemble defined by  $\mu_t$ reproduces the non-Markovian dynamics.\\

Trajectory simulations incorporating non-Markovian effects are initiated at $t=0$ with the system prepared in a selected initial state $\ket{\psi(0)}$, while the influence martingale is initialized as $\mu_0 \equiv \mu_{t=0} = 1$. 
For a given noise model defined by the set of operators $\{L_k\}$ and a specified choice of shifted positive rates $r_k(t)$, the infinitesimal evolution of the system state and the corresponding increment of the influence martingale over a time interval $dt$ are computed using Eqs.~\eqref{eq:itom2} and \eqref{eq:mart3}, respectively.
If a jump occurs in channel $k$ during the interval, i.e., $dN^{(k)}=1$, the martingale
increment $d\mu_t$ receives a discrete contribution from Eq.~\eqref{eq:mart5b}, 
in addition to the deterministic drift contribution from Eq.~\eqref{eq:mart5a}. 
If no jump occurs in channel $k$ during the interval, $dN^{(k)}=0$
and  $d\mu_t$ receives only  the drift contribution. 
At the subsequent time step, the system state and influence martingale updated from the preceding step are employed to compute the martingale increment via Eq.~\eqref{eq:mart3}, while the evolution of the state is propagated according to Eq.~\eqref{eq:itom2}. 
This recursive procedure is iterated until the final simulation time $t=T$ is reached.

\section{Extension to tensor network-based methods}
\label{sec:network}
In the MCWF method, unraveling reduces computational cost by evolving $d$-dimensional state vectors $\ket{\psi(t)}$ rather than the $d^2$-dimensional density matrices needed for direct  integration of GKSL master equation.   
Nevertheless, the state $\ket{\psi}$  of a system of $N$ subsystems ($i=1,2,\dots,N$) with local Hilbert space dimensions $d_i$,   spans the full Hilbert space of dimension $d = \prod_{i=1}^N d_i$, which grows exponentially with $N$. Consequently, the memory and computational cost for $N$-qubit ($d_i=2$) systems scale as $\mathcal{O}(2^N)$. Thus, although the MCWF method reduces the computational scaling from $\mathcal{O}(d^2)$ associated with density matrix evolution to $\mathcal{O}(d)$ for state vector propagation, it remains subject to the exponential complexity inherent to many-body quantum simulations \cite{weimersimulation2021}. 
Tensor network approaches  \cite{Werner2016} such as the Matrix Product State  \cite{perezgarciamatrix2007,RevModPhys.93.045003} and Matrix Product Operator  methods  \cite{PhysRevLett.93.207204} mitigate this exponential complexity by representing many-body quantum states as networks of lower-rank tensors connected through contracted indices. \\

\textbf{Matrix Product State   method:} In the MPS approach, a pure state $\ket{\psi}$ is represented as a one-dimensional chain of local tensors linked by contracted indices, with the \textit{bond dimension} $\chi$ controlling the maximum bipartite entanglement faithfully captured between subsystems \cite{perezgarciamatrix2007,RevModPhys.93.045003}. For an $N$-particle quantum system with   local Hilbert space dimension $d$, the full wave-function resides in a $d^N$-dimensional Hilbert space and can be written as the rank-$N$ tensor $c_{i_1 i_2 \cdots i_N}$ 
in the computational basis
as 
\begin{eqnarray}
{\k{\psi} = \sum_{i_1, i_2, \cdots i_N = 0}^{d-1} c_{i_1 i_2 \cdots i_N}
\k{i_1 i_2 \cdots i_N}.}
\label{eq:ss13}
\end{eqnarray}
The MPS  formalism represents a many-body quantum state $|\psi\rangle$ by decomposing the full coefficient tensor into a chain of site-dependent tensors (matrices) \cite{perezgarciamatrix2007}:
\begin{eqnarray}
|\psi\rangle = \sum_{i_1,i_2,\ldots,i_N = 0}^{d-1}  A^{[1]}_{i_1} A^{[2]}_{i_2} \cdots A^{[N]}_{i_N} |i_1 i_2 \cdots i_N\rangle\,,
\label{eq:ss14}
\end{eqnarray}
where each $A^{[k]}_{i_k}$ is a $\chi_{k-1} \times \chi_k$ matrix encoding the local state of site $k$. The index $i_k$, known as the `physical leg', spans the local Hilbert space of site $k$. Each site tensor $A^{[k]}_{i_k}$ also carries two additional indices, $a_{k-1} \leq \chi_{k-1}$ and $a_k \leq \chi_k$, referred to as the  left  and  right bond indices  or `virtual legs', which connect the site to its neighboring sites. Specifically, the left bond index links sites $k-1$ and $k$, while the right bond index links sites $k$ and $k+1$. The dimension ($\chi_k$) of a bond index  $a_k$, is termed the  bond dimension, and in practice, a fixed maximum bond dimension $\chi$ is often chosen such that $\chi_k \leq \chi$ for all $k$. The bond dimension $\chi$ sets an upper limit on the bipartite entanglement that the MPS can encode between the left subsystem $\{1,\dots,k\}$ and the right subsystem $\{k+1,\dots,N\}$. A small $\chi$ corresponds to a low-rank approximation with limited entanglement, whereas a larger $\chi$ allows for the representation of more highly entangled states. 
In a qubit-chain, $i_k$ takes values in ${0,1}$, corresponding to the basis states $\ket{0}$ (down) and $\ket{1}$ (up), so that each site is associated with two matrices, $A^{[k]}_{0}$ and $A^{[k]}_{1}$. 
However, for the simple product state $\ket{0,0,\ldots,0}$, all site tensors reduce to scalars, $A^{[k]}_{0} = 1$ and $A^{[k]}_{1} = 0$, indicating that, in the absence of inter-site correlations, the representation requires a bond dimension of unity.\\

An MPS can be expressed in different canonical forms depending on how orthogonality is imposed \cite{10.1063/1.5000784}. In the `left-canonical' form, the site tensors satisfy $\sum_{i_k} \left(A^{[k]}_{i_k}\right)^\dagger  A^{[k]}_{i_k}  = \mathbb{I}$,
so that each tensor acts as an isometry from left to right. In the `right-canonical' form, the tensors satisfy
$\sum_{i_k}  A^{[k]}_{i_k}  \left(A^{[k]}_{i_k}\right)^\dagger = \mathbb{I}$,
corresponding to isometries from right to left.  
When the state is not normalized or an operator is to be applied, 
it is convenient to represent the MPS in a mixed canonical form, in 
which an orthogonality center is chosen at a specific site $k$
with the {corresponding center tensor $M^{[k]}$, carries the overall norm for unnormalized states. Its singular values across a given bi-partition encode the Schmidt spectrum and hence the bipartite entanglement properties of the state.} This structure ensures numerical stability, 
allows for efficient manipulation of the MPS, and enables the 
straightforward computation of expectation values and Schmidt 
decompositions across any bipartition at the orthogonality center. \\
 
The MPS representation thus decomposes a many-body quantum state 
into a sequence of $N$ site tensors, each with one physical index 
of dimension $d$ and two virtual bond indices of dimension at most 
{$\chi = \chi_{max}$, containing approximately $d \times \chi^2$ parameters per tensor. Summing over all $N$ sites, the total number of parameters 
scales as $\mathcal{O}(N d \chi_{max}^2)$, thus compressing the storage requirements for a quantum state from the exponentially large $d^N$ to a more tractable scaling of $\mathcal{O}(N d \chi_{max}^2)$. However, the cost still grows polynomially with $\chi_{max}$, making the control of $\chi_{max}$ crucial for computational efficiency.}\\

\textbf{Matrix Product Operator  method:}
The matrix product operator (MPO) formalism--referred to as a
matrix product density operator (MPDO) when applied to 
density matrices--extends the  MPS  ansatz from pure states 
to mixed states (density operators) 
\cite{PhysRevResearch.3.023005}. 
Whereas MPS employs linear maps acting on the Hilbert space of
systems, MPO/MPDO generalizes this approach by applying completely positive (CP) maps on the operator space.
For an $N$-particle system with local dimension $d$, the MPDO represents the density matrix $\rho$ as a tensor network with the same  one-dimensional chain structure as an MPS, but with two physical indices per site corresponding to the bra and ket degrees of freedom: 
{
\begin{eqnarray}
\label{eq:ss15}
\rho &=& \sum_{i_1,j_1,\cdots, i_N,j_N = 0}^{d-1} \tr{\left(M_{i_1, j_1}^{[1]}M_{i_2, j_2}^{[2]}
\cdots M_{i_N, j_N}^{[N]}\right)} \nonumber \\
&& \qquad \cdot \quad \kb{i_1i_2\cdots i_N}{j_1 j_2 \cdots j_N}.
\end{eqnarray}
}
Each local tensor $M_{i_k,j_k}^{[k]}$ is a $\chi_{k-1}^2 \times \chi_k^2$ matrix (with $\chi_0=1$, $\chi_N=1$) and can be decomposed as{:
\begin{eqnarray}
M_{i_k, j_k}^{[k]}
&=&
\sum_{\alpha = 0}^{d_k-1} A_{i_k,\alpha}^{[k]} \otimes (A_{j_k,\alpha}^{[k]})^\star\,,
\label{eq:ss16}
\end{eqnarray}
}
where  for each pair $(i_k,\alpha)$, $A^{[k]}_{i_k,\alpha}$,
is a matrix of size
$\chi_{k-1}  \times \chi_k $ and $d_k$ is at most 
$d\chi_{k-1}\chi_k$. This decomposition ensures 
that each local tensor represents a completely positive map acting on the operator space.  
{Direct integration of GKSL master equation which evolves the full density matrix, is well suited to the MPDO framework, as MPDO  provides a compact representation of $\rho(t)$. For a maximum bond dimension $D = D_{max}$, the MPDO represents a density matrix of dimension $d^{2N}$ using only $\mathcal{O}(N d^2 D_{max}^2)$ parameters, yielding an exponential reduction in storage cost compared to the full $\mathcal{O}(d^{2N})$ scaling. The efficiency and accuracy of this representation is controlled by the choice of bond dimension $D_{max}$}, which reflects the amount of correlation or operator-space entanglement generated during the evolution.\\

In contrast, the MCWF method integrates efficiently with the MPS formalism, 
where each stochastic trajectory is encoded as an MPS rather than a full state vector in the exponentially large Hilbert space{, where deterministic non-Hermitian evolution between quantum jumps is carried out within the MPS manifold using algorithms such as TDVP \cite{Haegeman2011,PhysRevLett.100.130501} or TEBD \cite{Paeckel2019}, which approximate the dynamics within the low-entanglement variational manifold specified by the bond dimension $\chi$. Quantum jumps are typically implemented via local or two-body jump operators, followed by re-orthogonalization and, when necessary, truncation to maintain the chosen $\chi$. Since different stochastic trajectories are statistically independent, they can be propagated in parallel; however, the ensemble averaging required to reconstruct observable introduces a sampling overhead. Such schemes are particularly well-suited for simulating open quantum systems in regimes of restricted entanglement growth keeping the required bond dimension $\chi = \chi_{max}$ small without significant loss of accuracy for larger system sizes and longer evolution times. However, once entanglement growth or operator-induced correlations require a larger bond dimension, \(\chi_{\max}\) must be increased to maintain accuracy, leading to a corresponding increase in computational cost.}

\section{Algorithm of Tensor Jump Method}
\label{sec:tjm} 
The adaptation of the MCWF approach to 
tensor-network frameworks is realized through the Tensor Jump 
Method (TJM) \cite{sander2025largescalestochasticsimulationopen}, 
which embeds stochastic dynamics directly into MPS/MPDO 
representations and unifies unitary, dissipative, and stochastic 
processes within a Trotterized evolution. Quantum jumps are 
implemented as local tensor updates rather than global 
wave-function operations, thereby   
reducing computational overhead 
and enabling efficient simulation of open quantum systems.
The following outlines the step-by-step formulation of the  
TJM-algorithm for many-body open quantum systems.\\

\textbf{Initialization:} The simulation starts  from a  specified initial state $\ket{\psi(0)}$, encoded as an MPS, and evolved up to a final time $T = n\delta t$ in $n$ discrete steps of size $\delta t$. 
The initialization includes specifications 
of $n$, $\delta t$, the system Hamiltonian $H_0$ 
(encoded as an MPO, analogous to the MPDO representation in Eq.~\eqref{eq:ss15}), 
the set of single-site jump operators $L_k$ with coupling strengths $\sqrt{\gamma_k}$, 
and the maximum bond dimension $\chi$.\\

As discussed in Sec.~\ref{sec:unravelmark}, within the TJM framework the MCWF method evolves each pure-state trajectory between jumps under the effective non-Hermitian Hamiltonian 
$H_{\rm eff} = H_0 + H_D$, where  $H_0$ generates the coherent unitary dynamics and non-Hermitian drift term $H_D \equiv - \frac{i}{2} \sum_k \gamma_k L_k^\dagger L_k$  accounts for norm decay. 
At stochastic instants, the lost norm is restored by a quantum jump implemented via $L_k$ with probability proportional to 
$\langle \psi(t) | L_k^\dagger L_k | \psi(t) \rangle$, followed by re-normalization of the state. 
The TJM algorithm provides a seamless unification of unitary, dissipative, and stochastic processes, as outlined in Secs.~\ref{sec:tjm-trot}, \ref{sec:tjm-jump}, and \ref{sec:tjm-tdvp}, and discussed in detail in \cite{sander2025largescalestochasticsimulationopen}.

\subsection{Trotterized Evolution Incorporating Coherent and Dissipative Dynamics}
\label{sec:tjm-trot} 
 The non-unitary propagator $U_{\rm no-jump}(\delta t) \equiv e^{-iH_{\rm eff}\delta t}$ describes coherent and dissipative dynamics (excluding jumps) through the non-Hermitian operator $H_{\rm eff} = H_0 + H_D$, where the coherent ($H_0$) and dissipative ($H_D$) contributions do not commute. 
The Suzuki–Trotter decomposition \cite{Lubich2008} 
approximates $U_{\rm no-jump}(\delta t)$ by factorizing 
the exponential $e^{-i(H_0 + H_D)\delta t}$ of non-commuting 
operators into a product of exponentials of the individual 
operators, enabling efficient implementation. 
The resulting Trotterized propagator, prior to the inclusion of stochastic jumps, admits systematic expansions to achieve different orders of accuracy. 
For instance, at first and second order, the evolution is approximated respectively as 
\begin{eqnarray}
e^{-i(H_0 + H_D)\delta t} &=& e^{-i H_0 \delta t}e^{-i H_D \delta t} + \mathcal{O}(\delta t^2)   \label{eq:ss17} \\
\mbox{and}\quad e^{-i(H_0 + H_D)\delta t} &=& e^{-i H_D \delta t/2} e^{-i H_0 \delta t} e^{-i H_D \delta t/2} \nonumber \\
&& \qquad \qquad \qquad \quad + \mathcal{O}(\delta t^3). \label{eq:ss18}
\end{eqnarray}
Advancing from first- to second-order Trotterization improves the accuracy by reducing the time-step error from $\mathcal{O}(\delta t^2)$ to $\mathcal{O}(\delta t^3)$, while incurring only  a negligible increase in computational cost \cite{sander2025largescalestochasticsimulationopen}, as the additional operations involve a small number of local operator exponentials and the expensive MPO applications remain largely unchanged.
Following Eq.~\eqref{eq:ss18}, the  propagator
$U_{\rm no-jump}(\delta t)$  can be expressed as
\begin{eqnarray}
U_{\rm no-jump}(\delta t) &\approx& D(\delta t/2) \cdot U(\delta t) \cdot D(\delta t/2), \label{eq:ss19}
\end{eqnarray}
where $U(\delta t) \equiv e^{-i H_0 \delta t}$ denotes the unitary propagator of the coherent dynamics, and $D(\delta t) \equiv e^{-i H_D \delta t}$ represents the non-unitary contribution arising from dissipation.  
For a finite time $T = n\delta t$, the interval $[0,T]$ is partitioned into $n$ segments with endpoints $(0, \delta t, 2\delta t, \dots, n\delta t)$ and the  
 overall {non-unitary} no-jump propagator $U_{\rm no-jump}(T)$ is obtained by successive application of the single-step propagator $U_{\rm no-jump}(\delta t)$, giving
{
\begin{eqnarray}
	\label{eq:ss20}
	U_{\rm no-jump}(T)
	&=&
	\Big{[}  D(\delta t/2) \cdot U(\delta t) \cdot D(\delta t/2)\Big{]}^n \nonumber \\
	&=&
    D(\delta t/2)  U(\delta t) \cdot  \nonumber \\
    \label{eq:ss21}
    && \cdot \Big{(} D(\delta t) U(\delta t) \Big{)}^{n-1} 
   \cdot D(\delta t/2) . 
\end{eqnarray}
}
Within the second-order Trotterization scheme, the deterministic no-jump evolution described by Eq.\ \eqref{eq:ss19} 
introduces a half-step of dissipation before the unitary part.
Across successive steps, the intermediate half-dissipations combine into full \(D(\delta t)\) factors, leaving dissipation consistently shifted ahead of the unitary by $\delta t/2$. 
 This offset is compensated only at  the conclusion of the evolution, wherein the terminal factor $D(\tfrac{\delta t}{2})$   
  restores synchronization at $T = n \delta t$.  
 The structure of $U_{\rm no\mbox{-}jump}(T)$ (in Eq.~\eqref{eq:ss21}) admits the interpretation of an initial dissipative half-step $D(\tfrac{\delta t}{2})$, followed by $(n-1)$ iterations each with composite operation $\big(D(\delta t)U(\delta t)\big)$, and terminated with a final sequence $D(\tfrac{\delta t}{2})U(\delta t)$. 
 Accordingly, $U_{\rm no\mbox{-}jump}(T)$ can be expressed as a product of step-wise operators 
 $\{F^{\rm no-jump}_i(\delta t)\}_{i=0}^n$  \cite{sander2025largescalestochasticsimulationopen} 
{
\begin{eqnarray}
U_{\rm no-jump}(T) &=& \prod_{i=0}^n F^{\rm no-jump}_{n-i}(\delta t),
\label{eq:ss21a}
\end{eqnarray}
}
where
{
\begin{eqnarray}
\label{eq:ss21b}
F^{\rm no-jump}_i(\delta t) 
& \equiv &  
\begin{cases}
D(\delta t/2) \quad \mbox{for } i = 0 \\ 
D(\delta t) U(\delta t)\quad \mbox{for } 0<i<n\\
D(\delta t/2)  U(\delta t) \quad \mbox{for } i = n 
\end{cases}\,.
\end{eqnarray}
}
The evolved state $U_{\rm no\mbox{-}jump}(T)\ket{\Psi(0)}$ then represents the system at time $T$, corresponding to a single quantum trajectory in which no jump event occurs during the interval $t \in [0,T]$. \\

 In a lattice of $N$ sites with local Hilbert space dimension $d$, each jump operator acts nontrivially  on a specific site $\ell$, taking the form $\mathbb{I}^{(\ell-1)} \otimes L_k^{[\ell]} \otimes \mathbb{I}^{(N-\ell)}$
with the total contribution obtained by summing over all sites $\ell = 1, \dots, N$.
This structure enables the dissipative Hamiltonian, 
$H_D = -\tfrac{i}{2}\sum_{k}  \gamma_k L_k^\dagger L_k,$
to be expressed as a sum of site-local contributions,
$$
 H_D = -\tfrac{i}{2}\sum_{\ell=1}^N \sum_{k \in S(\ell)} 
 \gamma_k 
 \left[ \mathbb{I}^{(\ell - 1)} \otimes  
 \left( {L_k^{[\ell]} }^\dagger L_k^{[\ell]}  \right)
 \otimes \mathbb{I}^{(N-\ell)} \right]
$$
where $S_{\ell}$   is the set of all jump operators acting on site $\ell$. Since the site-local terms commute for two different 
sites, the dissipative evolution operator factorizes as 
{
\begin{eqnarray}
D(\delta t) &=& \bigotimes_{\ell=1}^N D_\ell(\delta t) \nonumber \\
\label{eq:ss22}
\quad \mbox{with, } 
\quad D_\ell[\delta t] &=& \exp\!\Big[-\tfrac{\delta t}{2}\sum_{k\in S_\ell} \gamma_k (L_k^{[\ell]})^\dagger L_k^{[\ell]}\Big]\,.
\end{eqnarray}
}

The dissipative dynamics therefore reduces to a product of local 
$d\times d$  maps, a structure naturally compatible with tensor-network implementations.

 \subsection{Integration of Quantum Jumps into Trotterized No-Jump Dynamics} 
 \label{sec:tjm-jump}
 The finite-time no-jump propagator 
$U_{\rm no-jump}(T)$ 
governs the deterministic evolution of an initial state 
 $\ket{\Psi(0)}$ up to time $T$ under the combined unitary 
and continuous dissipative dynamics, while explicitly excluding 
stochastic jump events.
An effective trajectory-wise propagator $\overline{U}(T)$, which 
consistently accounts for unitary, dissipative, and stochastic jump 
dynamics, is constructed by augmenting the {non-unitary} no-jump propagator 
$U_{\rm no-jump}(T)$ from Eq.~\eqref{eq:ss21a} with stochastic jump 
operators $J_j(\delta t, \varepsilon)$ at each step $j$, 
following the application of $F_j^{\rm no-jump}(\delta t)$. 
The resulting{, full trajectory-wise} propagator $\overline{U}(T)$ thus encodes the combined unitary, dissipative, and stochastic jump dynamics over the interval $t \in [0,T]$, and can be expressed as 
\begin{eqnarray}
\overline{U}(T) =  \prod_{j=0}^n J_{n-j}(\delta t,\varepsilon) F^{\rm no-jump}_{n-j}(\delta t) = \prod_{j=0}^n    F_{n-j}(\delta t), \nonumber \\
\label{eq:ss22a}
\end{eqnarray}
where $F_j(\delta t) \equiv J_j(\delta t,\varepsilon) F^{\rm no-jump}_{j}(\delta t)$.\\

In the second-order Trotter scheme, dissipative evolution is applied as a half-step before the unitary propagation, with the compensating half-step deferred to the final stage. This prevents direct access to the physical state $\ket{\psi(j\delta t)}$ at intermediate sampling times $(0, \delta t, 2\delta t, \dots, n\delta t)$, as the final correction $F_n(\delta t)$ has not yet been applied.
To enable trajectory sampling at all discrete times without compromising higher-order Trotter accuracy, we introduce an auxiliary MPS, $\ket{\Phi(j\delta t)}$, whose evolution is defined recursively through 
{
\begin{eqnarray}
\k{\Phi(\delta t)} &=& F_0(\delta t)\,\k{\psi(0)}, \nonumber \\
\k{\Phi((j+1)\delta t)} &=& F_j(\delta t)\, \k{\Phi(j\delta t)}, \, (0<j<n)
\label{eq:sampling}
\end{eqnarray}
}

allowing $\ket{\Phi(j\delta t)}$ to act as an intermediate carrier of the dynamics. The physical state can then be recovered at any time step by applying the final correction,
$
\k{\psi(j\delta t)} = F_n(\delta t)\,\k{\Phi(j\delta t)}
$. 
Each $F_j(\delta t)$ in Eq.~\eqref{eq:sampling} contains a factor $J_j(\delta t,\varepsilon)$ that encodes the stochastic jump process occurring within the interval $[j\delta t, (j+1)\delta t]$. Its action is defined by applying $J(\delta t,\varepsilon)$ to the intermediate state $F^{\rm no\text{-}jump}_j(\delta t)\ket{\Phi(j\delta t)}$, yielding the updated state $\ket{\Phi((j+1)\delta t)}$ after proper normalization.  
The probability that the  intermediate state  undergoes a jump is given by 
\begin{eqnarray}
    \delta p = 1 -
    \bra{\Phi(j\delta t)} 
    (F^{\rm no\text{-}jump}_j)^\dagger F^{\rm no\text{-}jump}_j 
    \ket{\Phi(j\delta t)}.
\label{eq:ss26} 
\end{eqnarray}
To determine whether a jump occurs, a random number 
$\epsilon \in [0,1]$ is drawn from a uniform distribution.  
If $\epsilon \geq \delta p$, no jump occurs and $J(\delta t,\epsilon)$ is 
identified with the identity operator and the state propagated to the next time-step iteration is
{
\begin{eqnarray}
    \ket{\Phi((j+1)\delta t)} 
    =  
    F^{\rm no\text{-}jump}_j(\delta t)\ket{\Phi(j\delta t)}.
\label{eq:ss27}
\end{eqnarray}
}

If instead $\varepsilon < \delta p$,
a stochastic jump takes place through one of the 
available channels $\{k\}_{k=1}^{K_c}$, where $K_c$ denotes the total number of channels.  
The jump channel is selected by sampling the index $k \in \{1,\dots,K_c\}$ 
according to the probability distribution $\{\delta p_k/\delta p\}_{k=1}^{K_c}$, with 
{
\begin{eqnarray}
    \delta p_k = \gamma_k \, \delta t \, 
    \bra{\Phi(j\delta t)} 
    (F^{\rm no\text{-}jump}_j)^\dagger L_k^\dagger L_k F^{\rm no\text{-}jump}_j 
    \ket{\Phi(j\delta t)}, \nonumber \\
\label{eq:ss28}
\end{eqnarray}
}
defining the probability of a jump through channel $k$.  
After the jump occurs in channel $k$, the state is updated and re-normalized as  
\begin{eqnarray}
\ket{\Phi((j+1)\delta t)} 
= \frac{\sqrt{\gamma_k \delta t}}{\sqrt{\delta p_k }} 
L_k F^{\rm no\text{-}jump}_j(\delta t)\ket{\Phi(j\delta t)},
 \label{eq:ss29}
\end{eqnarray}
and subsequently propagated to the next iteration of the simulation.

\subsection{Implementing Unitary Evolution via Dynamic TDVP in Tensor Networks}
\label{sec:tjm-tdvp} 
The unitary  factor $U(\delta t) \equiv e^{-iH_0\delta t}$ in the stepwise operators $\{F_j(\delta t)\}_{j=1}^n$ is
implemented using a dynamic   TDVP  framework which
restricts unitary dynamics to the   MPS manifold 
by applying the projector $P_{M,\ket{\psi(t)}}$
that by projects the
\sch \ equation onto the tangent space at the current state $\ket{\psi(t)}$.  The resulting projected evolution equation \cite{Haegeman2013},
\begin{eqnarray}
\frac{d}{dt}\ket{\psi(t)} = -i P_{M,\ket{\psi(t)}} H_0 \ket{\psi(t)}\,,
\label{eq:tdvp1}
\end{eqnarray}
provides the variationally optimal approximation to the exact dynamics
at fixed bond dimension,  
ensuring that the MPS structure is preserved throughout the evolution.\\

The dynamical TDVP framework \cite{Haegeman2011,PhysRevLett.100.130501} adopt a hybrid strategy combining two-site 
TDVP (2-TDVP) and one-site TDVP (1-TDVP).
In the 1-TDVP scheme, the
 Hamiltonian $H_0$ is represented as an MPO,
 while  the state $\k{\psi}$ is expressed in mixed-canonical 
 MPS form 
 with the orthogonality center at site $k$.
The local update is then performed on the site tensor
$M^{[k]}$ at the orthogonality center. 
 Its evolution is given by  
{
\begin{eqnarray}
\frac{dM^{[k]}(t)}{dt} = -i H_{0,\mathrm{eff}}^{[k]} M^{[k]}(t),
\label{eq:tdvp2}
\end{eqnarray}
}
where  $H_{0,\mathrm{eff}}^{[k]}$
is the effective single-site Hamiltonian acting on site $k$,   obtained by contracting the  MPO  representation of the global Hamiltonian $H_0$ with the left and right MPS environments. 
The time evolution proceeds via symmetric forward and backward sweeps:  during the forward sweep, site tensors are updated sequentially from the left boundary to the right ($k = 1,\cdots, N$), each evolved for a duration $\delta t/2$. After every local update, the orthogonality center is shifted to the next site via a QR decomposition  preserving the mixed-canonical structure of the MPS \cite{Haegeman2011,PhysRevLett.100.130501}.
Once the right boundary is reached, a backward sweep is performed in which the tensors are evolved from right to left ($k = N,\cdots, 1$) for another time interval $\delta t/2$, with the orthogonality center shifted using an LQ decomposition. A complete time step thus consists of a forward and backward sweep, yielding a symmetric integrator with second-order accuracy in the step size
$\delta t$. 
Since the evolution is entirely constrained within the original tangent space of the MPS manifold, the bond dimension remains fixed throughout the 1-TDVP simulation. 
In the 2-TDVP scheme, starting from an MPS in mixed-canonical form with the orthogonality center at site $k$,
the local MPS block is written as the tensor product 
$M^{[k,k+1]} \equiv M^{[k]} \otimes M^{[k+1]}$ of the  
pairs of neighboring site tensors
obtained by contracting the bond index between sites $k$ and $k+1$,
which is then 
  evolved under the two-site effective Hamiltonian $H_{0,\mbox{eff}}^{[k,k+1]}$ as
{
\begin{eqnarray}
\frac{d}{dt} M^{[k,k+1]}(t)   &=& -i H_{0,\mathrm{eff}}^{[k,k+1]} M^{[k,k+1]}(t), 
\label{eq:tdvp3}
\end{eqnarray}
}
followed by a singular value decomposition $M^{[k,k+1]} = USV^\dagger$ with truncation applied to $S$  when necessary to control the bond dimension. 
This approach allows the MPS to dynamically increase its bond dimension.\\

In the initial stages, the dynamic TDVP is carried out using the  2-TDVP  scheme, which permits adaptive bond-dimension growth and captures the development of entanglement during early evolution. Once the bond dimension reaches a predetermined maximum {$\chi = \chi_{\max}$}, the simulation 
transitions to the  1-TDVP  scheme, wherein the time evolution is restricted to the fixed variational manifold defined by this maximum bond dimension.

{
\subsection{Computational complexity: }
\label{sec:ccomplexity}
Ref.~\cite{sander2025largescalestochasticsimulationopen} shows that the computational cost of the tensor jump method (TJM) is dominated by TDVP sweeps. For a system with $N$ sites, local Hilbert-space dimension $d$, and maximum MPS bond dimension $\chi_{\max}$, Hamiltonian MPO bond dimension $D_H$, for $N_{traj}$ trajectories evolved over $n$ time steps, the overall dominant computational cost scales as $\mathcal{O}(n\, N_{traj}\, N\, \chi_{\max}^3\,  [d^2 + dD_H])$, assuming a fixed number of local jump operators per site, rare jump events, and $d,D_H\ll \chi_{\max}$. Thus within tolerable error limit, this approach allows enough bond dimension to capture entanglement growth for long time evolution. This advantage is however model dependent. Rapid entanglement growth may require a larger MPS bond dimension $\chi_{\max}$, while complex or highly non-local Hamiltonian may require a larger MPO bond dimension $D_H$, increasing the computational cost required to achieve the same accuracy.\\

For comparison, direct Lindblad evolution using a matrix product operator (MPO) representation of the density matrix scales as $\mathcal{O}(n\, N\, d^4 D^2 D_H^2)$, where $D$ is the MPO bond dimension. Since representing mixed states typically requires significantly larger bond dimensions ($D \gg \chi_{\max}$), trajectory-based MPS approaches are generally substantially more efficient for comparable accuracy. A detailed numerical comparison between MPS-based stochastic trajectory propagation and direct MPO-based Lindblad evolution has been reported in Ref.~\cite{sander2025largescalestochasticsimulationopen}. \\

In the present work, the martingale estimator introduces only a small additional overhead. Since only one martingale value is stored and updated once per trajectory at each time step, the storage and computational costs associated with the martingale updates scale as $$\mathcal{O}(N_{traj})\, \text{ and, }\,\mathcal{O}(n N_{traj})\quad \text{respectively}.$$ Consequently, this overhead is negligible compared to the dominant cost of the MPS trajectory propagation discussed in Ref.~\cite{sander2025largescalestochasticsimulationopen}. \\

}

\section{Numerical Implementation, Results and Discussions}  
\label{sec:nu}
 To illustrate the unraveling of Markovian and non-Markovian dynamics in a many-body setting, we consider the prototypical one-dimensional spin-$\tfrac{1}{2}$ transverse-field Ising chain, characterized by nearest-neighbor Ising interactions along the $z$-axis and subjected to a uniform transverse magnetic field along the $x$-axis. 
The corresponding Hamiltonian is 
\begin{eqnarray}
H_0 
&=& 
-J \sum_{i=1}^{N-1} \left(    Z^{[i]}  Z^{[i+1]} \right)
- g \sum_{i=1}^{N} X^{[i]} \,,
\label{eq:nu1}
\end{eqnarray}
where $X^{[i]}, Z^{[i]}$ denote Pauli $X$ and $Z$ operators acting on site $i$.
The first term, proportional to $J$, represents nearest-neighbor 
 exchange interactions, while the second term 
describes the coupling to a homogeneous magnetic field of strength $g$ 
along the $x$-axis.
This model serves as a canonical testbed for investigating 
correlated quantum dynamics: it is sufficiently simple to permit analytical 
treatments, yet rich enough to capture 
 the essential features of interacting quantum many-body systems,  
and is therefore widely employed as a standard reference 
across diverse domains of quantum physics. 
 For the numerical simulations presented here, we set $J=1$  
and 
$g=0.5$, making the magnetic field half as strong 
as the spin--spin interaction.
The Hamiltonian\ \eqref{eq:nu1} is first encoded in MPO form and 
fed into the TDVP engine 
to compute the unitary evolution 
of the MPS for each trajectory at every 
time step, prior to the application of dissipative and jump processes.\\

The fundamental noise modes arising from system–environment interactions are described by a set of site-local Lindblad operators $\{L_k\}$, where each operator corresponds to a specific noise channel labeled by $k$. In this work, we focus on three noise models that are particularly relevant for open qubit-chain dynamics: the dephasing, excitation, and relaxation channels, respectively represented by
\begin{eqnarray}
\label{eq:channels}
&&L_{\text{deph}} = Z =  \begin{pmatrix} 1 & 0 \\ 0 & -1 \end{pmatrix}, \nonumber \\
L_{\text{exc}} = \sigma^+ &=& \begin{pmatrix} 0 & 0 \\ 1 & 0 \end{pmatrix}, \quad
L_{\text{rel}} = \sigma^- = \begin{pmatrix} 0 & 1 \\ 0 & 0 \end{pmatrix}\,.
\end{eqnarray}
The dephasing channel, governed by the Pauli-$Z$ operator, 
randomizes the relative phase
between spin-up and spin-down states, 
thereby diminishing phase coherence. 
The excitation channel, governed by the 
raising operator $\sigma^+$, captures 
environment-driven transitions from the ground state (spin-down) to the excited state (spin-up), while the relaxation channel, represented by the lowering operator $\sigma^-$, accounts for spontaneous decay from the excited state back to the ground state. 
Collectively, these noise processes encapsulate the principal mechanisms of decoherence that dictate coherence times and mediate energy exchange in realistic qubit-chain systems.  
In our numerical simulations, the operators are scaled with appropriate normalization factors to satisfy the completeness condition $\sum L_l^\dagger  L_l = \mathbb{I}$ where the sum runs over {all channels per site, providing a convenient normalization for the auxiliary martingale trajectory process and for the evaluation of the shifted jump intensities.}\\

 {Decay rates in open quantum systems arise as theoretical quantities fixed by the
 	underlying microscopic system--environment dynamics \cite{Breuer2016, donvil2022}. In structured reservoir models, the spectral density determines the bath correlation function, which encodes non-Markovian memory effect in the reduced dynamics\cite{Breuer2016}. From the resulting dynamical map or propagator, one constructs a time-local generator that can be expressed in a generalized Lindblad form. The decay rates are then obtained as the eigenvalues of the decoherence matrix, associated with the dissipative part of the time-local generator\cite{4c01632}. For trajectory simulations exhibiting non-Markovian features, these decay rates ${\gamma_k(t)}$ are generally time-dependent and may become negative.} we adopt a benchmark model with damped oscillatory rate{\cite{donvil2022}},
\begin{eqnarray}
\gamma_k(t) = \gamma_\infty - B e^{-f(t)}\sin(\omega t),
\label{eq:chosengamma}
\end{eqnarray}
{which provides a prototypical description which mimics the damped oscillatory rates, arising from structured reservoirs with Lorentzian spectral densities, as obtained in time-local/TCL master-equation treatments \cite{PhysRevA.81.052103}. These models naturally yield exponentially damped oscillatory behavior and temporary negativity of decay rates, reflecting possible finite reservoir memory and non-Markovian information backflow.
The parameter $\gamma_\infty$ specifies the asymptotic Markovian 
background dissipation, representing the residual steady decay 
rate in the long-time limit ($t \to \infty$). The oscillatory
component $B e^{-f(t)}\sin(\omega t)$ are commonly associated with memory-induced 
feedback from a structured environment, with $\omega$ denoting the 
characteristic frequency of the dominant environmental mode. The 
damping factor $e^{-f(t)}$ (for $f(t) > 0$) imposes a finite 
correlation time of the bath, ensuring the gradual suppression of 
memory effects. For numerical simulations, we choose a parameter set $f(t) = 0.25 t^3$, $\omega = 7.5$, {$B = 0.2$, and $
\gamma_\infty \simeq 0.137 $, motivated from \cite{PhysRevA.81.052103}}. This parameter set yields finite time intervals in which $\gamma_k(t)<0$, and
therefore provides a useful phenomenological benchmark for testing the trajectory method in a controlled non-Markovian regime.
}

\begin{figure}[!h]
\centering
\includegraphics[width=0.95\linewidth]{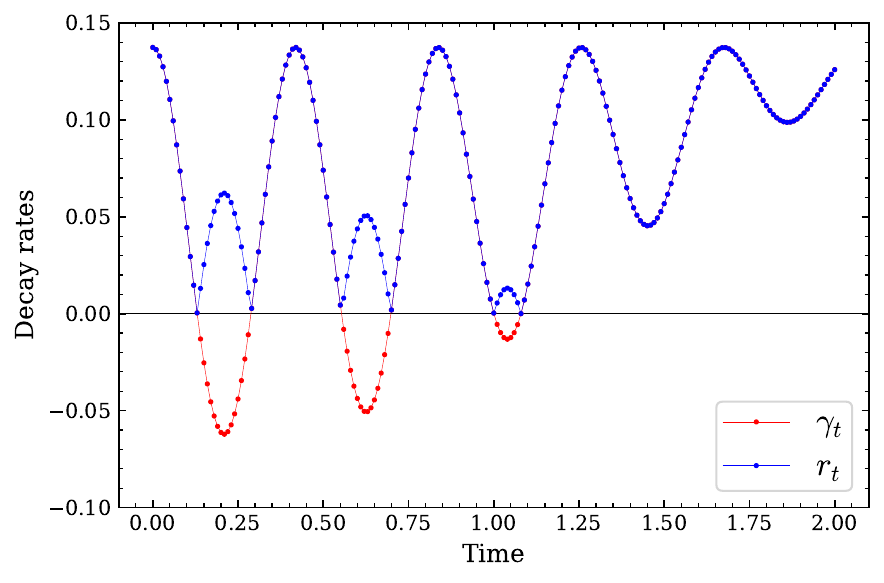}
\caption{Plot of decay rate $ \gamma_k(t)$ of the form Eq.~\eqref{eq:chosengamma} with chosen parameter set $f(t) = 0.25 t^3$, $\omega = 7.5$, $B = 0.2$, and $\gamma_\infty \simeq 0.137 $, and shifted Decay rate $r_k(t)$ vs time}
\label{fig:shift}
\end{figure}

Whenever $\gamma_k(t)$ becomes negative, we introduce a time-dependent shift $C_t$ such that the shifted rates $r_k(t) = \gamma_k(t) + C_t$ remain {positive}, ensuring consistency with martingale-based computations. In our simulations,
 $C_t$ is set to zero when all $\gamma_k(t)$ are positive, and otherwise defined as twice the modulus of the most negative rate \cite{donvil2022} , i.e.
\begin{eqnarray}
C_t = -2 \min \{0, \gamma_1(t), \gamma_2(t), \ldots\},
\label{eq:cchoice}
\end{eqnarray}
which guarantees positivity of $r_k(t)$ for all channels and at all time. Fig.~\ref{fig:shift} illustrates the temporal profiles of $\gamma_k(t)$ from Eq.~\eqref{eq:chosengamma} together with the corresponding shifted rates $r_k(t)$ 
  obtained using the shift $C_t$ specified in Eq.~\eqref{eq:cchoice}{, where $\gamma_1(t), \gamma_2(t), \ldots $ are decay rates with temporarily negative values, either taken for all the sites or sites within the `Influence radius' as discussed in Sec.~\ref{sec:infrad}. As this does not explicitly depend on the dynamics, it is pre-calculated for numerical convenience. Fig.~\ref{fig:shift} shows the decay rate $\gamma_k(t)$ and shifted decay rate $r_k(t)$ for the chosen parameters. }\\
  
We consider a spin-chain of length $N$ initialized in the pure product state with all spins in the {ground state} configuration 
$\ket{\psi(0)} = \ket{0}^{\otimes N}$ and simulate the non-Markovian dynamics, with a {maximum bond dimension $\chi_{max}$.}
Starting from the  state $\ket{\psi(0)} = \ket{0}^{\otimes N}$, trajectory simulations of the non-Markovian dynamics are carried out as described in Sec.~\ref{sec:unravelnonmark}. The stochastic evolution is governed by the shifted decay rates $r_k(t)$, with the conditional expectation
$
\mathbb{E}\!\left[ dN_t^{(k)} \,\big|\, |\psi(t)\rangle \right] 
= r_k(t)\, \| L_k |\psi(t)\rangle \|^2 \, dt$
defining the probability of a quantum jump within $[t,t+dt]$ for {$k$-th} channel, to first order in \(dt\).\\

\begin{figure*}[t]
    \centering
    \begin{subfigure}{0.48\linewidth}
        \centering
        \includegraphics[width=\linewidth]{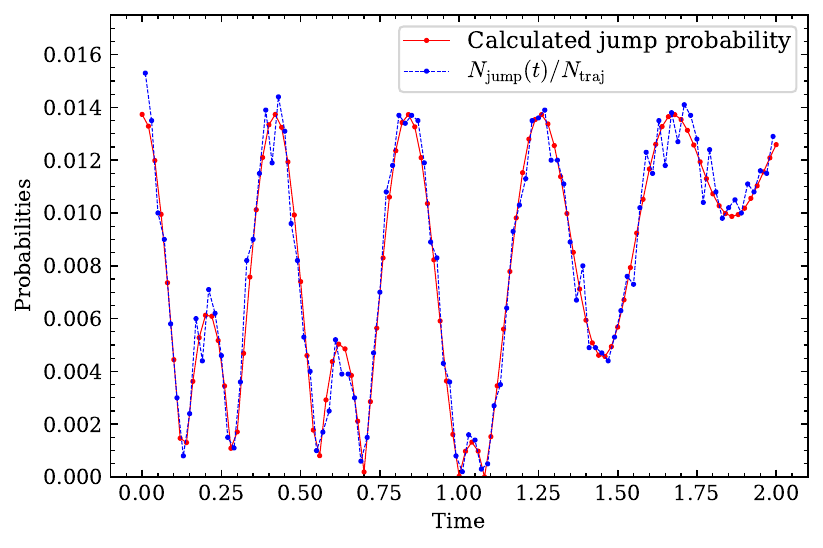}
    \end{subfigure}%
    \hfill
    \begin{subfigure}{0.495\linewidth}
        \centering
        \includegraphics[width=\linewidth]{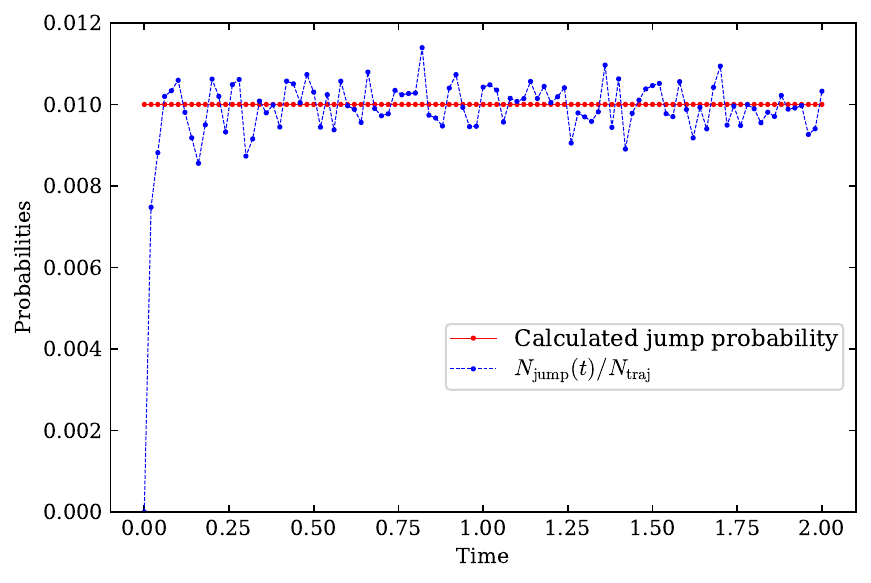}
    \end{subfigure}
    
    \caption{{Time dependence of the jump probability under dephasing noise at all sites. (a) non-Markovian dynamics of a spin-chain with $N=5$, simulated using the decay rates in Fig.~\ref{fig:shift}, with $dt=0.02$, $\chi_{\max}=4$, and $N_{\mathrm{traj}}=1000$ trajectories. Here, $N_{\mathrm{jump}}(t)/N_{\mathrm{traj}}$ is the fraction of trajectories exhibiting a jump in the interval $[t,t+dt]$ and is compared with the calculated jump probability $\sum_k r_k(t)\|L_k|\psi(t)\rangle\|^2 dt$. (b) Corresponding Markovian case for the same dephasing channel at all sites with constant positive decay rates $\gamma_k=0.1$.}}
	\label{fig:Calcjactj}
	\end{figure*}

From a simulation of {a spin-chain with $N = 5$ subject to dephasing noise with decay rates of Fig.~\ref{fig:shift}, with $dt=0.02$, $\chi_{max} = 4$, } $N_{\mathrm{traj}}=1000$  trajectories, we record the number of  trajectories $N_{\mathrm{jump}}(t)$ in which a jump occurred during $[t,t+dt]$. 
The ratio $N_{\mathrm{jump}}(t)/N_{\mathrm{traj}}$, {along with the calculated jump probability $\sum_k r_k(t)\|L_k|\psi(t)\rangle\|^2 dt$, which provides an empirical estimate of the probability in the limit of large $N_{\mathrm{traj}}$}, is shown in the panel (a) of Fig.~\ref{fig:Calcjactj} for the dephasing channel with . For comparison, the Markovian case with a constant positive decay {rate $\gamma_k = 0.1$} is presented in the panel (b) of Fig.\ \ref{fig:Calcjactj}.\\

\begin{figure}[t]
    \centering
    \includegraphics[width=0.95\linewidth]{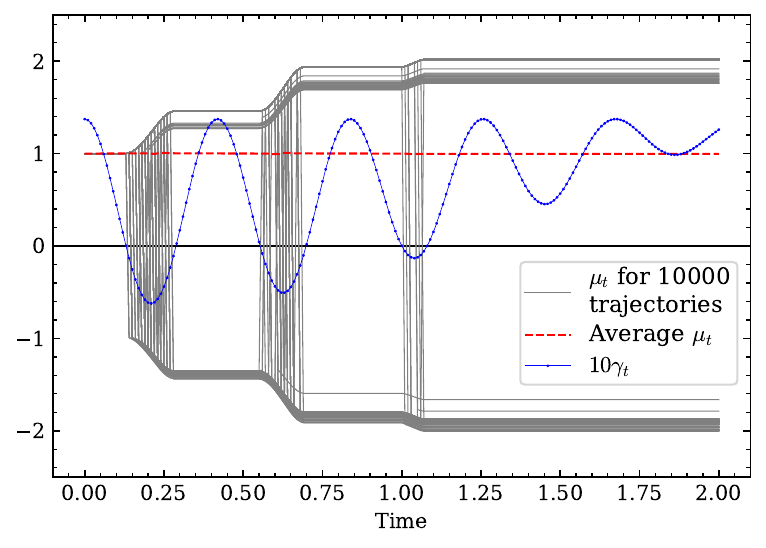}
\caption{Time evolution of the martingale factor $\mu_t$ for all stochastic trajectories under non-Markovian dephasing noise. The {red} horizontal line indicates the ensemble-averaged value of $\mu_t$ across all trajectories. {$10 \cdot\gamma(t)$ } overlaid to illustrate the martingale behavior during intervals where $\gamma(t)$ attains negative values.}
\label{fig:marttraj}
\end{figure}

The evolution of the influence martingale $\mu_t$ is computed from 
Eq.~\eqref{eq:mart3} with the initial condition $\mu_0=1$. 
Fig.~\ref{fig:marttraj} shows the time profiles of $\mu_t$ for {spin-chain length $N = 30$, under $L_{\rm{deph}}$, simulated with $N_{\mathrm{traj}}=10,000$ trajectories, maximum bond dimension $\chi = 4$} together with their ensemble 
average {$\mathbb{E}[\mu_t]$ remains close to unity}. Comparison with the time profile of the unshifted decay rate $\gamma_k(t)$ {of Fig.~\ref{fig:shift} scaled by 10}, also displayed in the same figure, reveals that $\mu_t$ exhibits temporal variations during intervals where $\gamma_k(t)$ takes negative values, while remaining constant when $\gamma_k(t)$ is positive. 
This behavior demonstrates that the shifts in decay rates, introduced during intervals of negative $\gamma_k(t)$ to ensure a consistent probabilistic interpretation, manifest as {time-dependent changes in $\mu_t$.}
\subsection{Benchmarking:}
\label{sec:bmrkn}
The expectation value of an observable $\hat{O}$ in the state $\rho(t)$ is given by 
$\mathrm{Tr}(\hat{O}\rho)$, which can equivalently be expressed as an {normalized ensemble average,
$\langle \hat{O} \rangle_t = \frac{\mathbb{E}\!\left[\mu_t \,
\langle \psi(t) | \hat{O} | \psi(t) \rangle \right]}{\mathbb{E}[\mu_t]}$}, 
where the average is taken over all stochastic trajectories at time $t$, weighted by the 
corresponding martingale factor $\mu_t$. {In the infinite-trajectory limit \(\mathbb{E}[\mu_t]=1\), while the explicit normalization is retained in finite samples to compensate or statistical fluctuations of the martingale weights.} We simulate the temporal evolution of the local observable $O_i = X^{[i]}$ {at site $i$,} in the transverse-field Ising chain. In the absence of noise, the coherent dynamics of the closed system are governed by the Hamiltonian $H_0$ (Eq.~\eqref{eq:nu1}), where the $ZZ$ term favors ferromagnetic alignment of neighboring spins along the $z$-axis, while the transverse-field 
term involving $X$ induces spin polarization along the $x$-axis through quantum spin flips. The expectation value $\langle O_i \rangle = \langle X^{[i]} \rangle$ thus quantifies the local transverse magnetization arising from the competition between the nearest-neighbor interaction-induced ordering along the $z$-direction and the transverse-field-driven spin alignment along the $x$-axis. The coherent 
dynamics generated by $H_0$ are modified in the presence of environmental noise, where different noise operators $L_k$ lead to distinct dissipative effects on the system’s evolution. \\

\begin{figure*}[t]
	\centering
	\begin{subfigure}{0.48\linewidth}
		\centering
		\includegraphics[width=\linewidth]{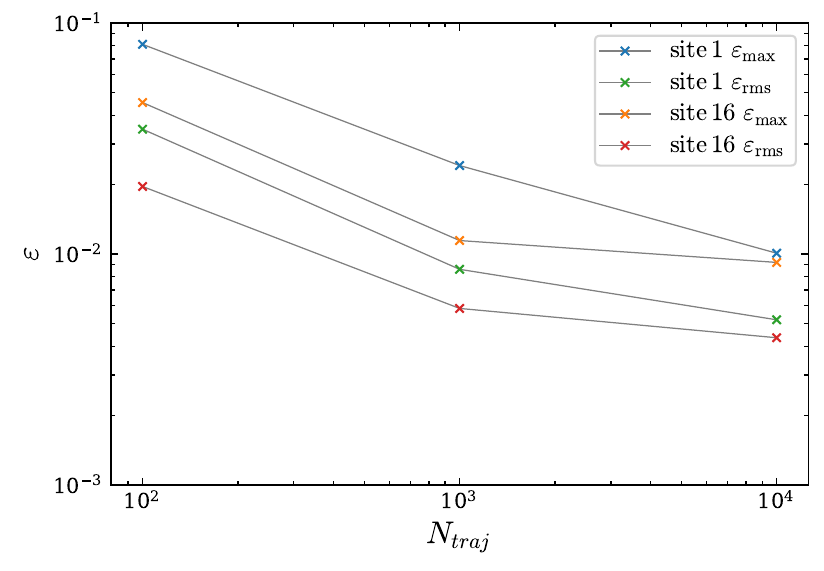}
	\end{subfigure}%
	\hfill
	\begin{subfigure}{0.5\linewidth}
		\centering
		\includegraphics[width=\linewidth]{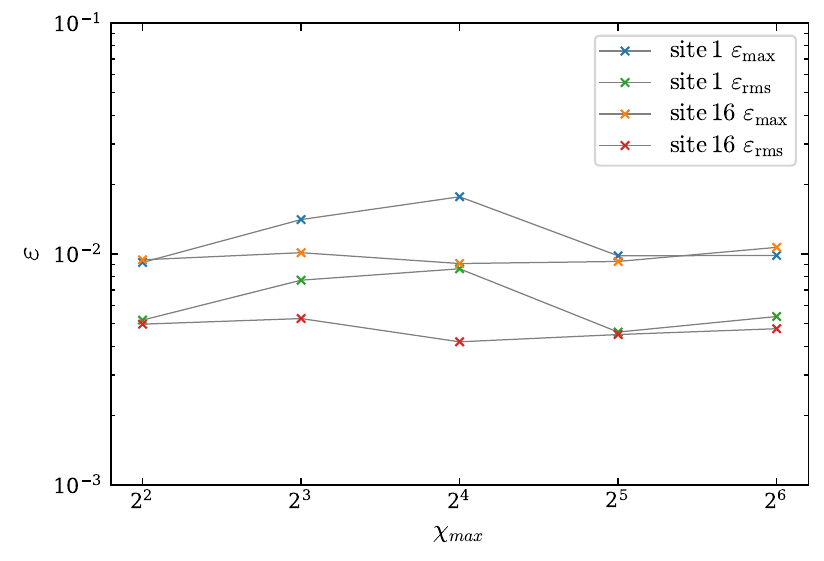}
	\end{subfigure}
	\caption{{$\varepsilon_{\mathrm{rms}}$ and $\varepsilon_{\max}$ (a) with the number of trajectories $N_{\rm traj}$ at fixed $\chi_{\max}=128$, showing systematic Monte Carlo convergence.
			(b) with the maximum bond dimension $\chi_{\max}$ at fixed
			$N_{\rm traj}=10\,000$, showing that the simulation is already converged with respect to bond dimension for the parameters considered. Errors are computed for the normalized observables at
			sites $i=1$ and $i=16$ relative to the numerically exact MPO-based solution, for a 30-site chain subject to $L_{\rm{deph}}$ with decay rates $\gamma(t)$ of Fig.~\ref{fig:shift} applied uniformly to all sites. }}
	\label{fig:convergence}
\end{figure*}

{ 
	For benchmarking we take the dephasing noise operator ($L_{\text{deph}}$) in \eqref{eq:channels}. For all cases the time step is chosen to be $dt = 0.01$, and the time evolution is calculated for total time $T = 2$.
	
	We assess the accuracy of the trajectory-based calculation by comparing the
	trajectory-averaged expectation value $\langle O_i \rangle = \langle X^{[i]} \rangle$ at site $i$, with benchmark results obtained from numerically exact MPO-based simulations performed with a customized implementation of the \texttt{LindbladMPO} package introduced in \cite{SciPostLindbladMPO}, extended here to accommodate time-dependent decay rates $\gamma(t)$. The deviation is quantified using both
	the root-mean-square error over $n$ time steps,
	{
	\begin{equation}
		\varepsilon_{\mathrm{rms}}
		=
		\sqrt{
			\frac{1}{n}
			\sum_{m=1}^{n}
			\left[
			\langle O_i(t_m)\rangle_{\rm traj}
			-
			\langle O_i(t_m)\rangle_{\rm MPO}
			\right]^2
		},
	\end{equation}
}
	and the maximum error,
	{
	\begin{equation}
	 \varepsilon_{\max}
		=
		\max_{t_m}
		\left|
		\langle O_i(t_m)\rangle_{\rm traj}
		-
		\langle O_i(t_m)\rangle_{\rm MPO}
		\right|.
	\end{equation}
}
	These error measures allow us to analyze convergence with respect to both the
	number of trajectories $N_{\rm traj}$ and the maximum MPS bond dimension $\chi_{\max}$, which are fixed before a given simulation run. In the absence of MPO-based exact methods, this value can also be determined through convergence tests. Specifically, one may systematically increase \(\chi_{\max}\) or \(N_{\mathrm{traj}}\) until the expectation values of interest become insensitive to further increases within the desired accuracy.\\

	\textbf{Monte Carlo convergence:} We first examine convergence with respect to the number of trajectories $N_{\rm traj}$, while fixing the maximum bond dimension $\chi_{\max} = 128$. Fig.~\ref{fig:convergence}(a) shows the root-mean-square error, \(\varepsilon_{\mathrm{rms}}\), and the maximum error, \(\varepsilon_{\max}\), evaluated at sites \(1\) and \(16\) for a system size of \(30\) sites subject to $L_{\rm{deph}}$ and decay rates same as Fig.~\ref{fig:shift} at all sites. As the number of trajectories increases, both error measures decrease significantly, indicating convergence of the trajectory-averaged results.\\

	\textbf{Convergence over Bond dimension:} We next examine convergence with respect to the maximum bond dimension $\chi_{\max}$, keeping the number of trajectories fixed at \(N_{\mathrm{traj}}=10000\). Fig.~\ref{fig:convergence}(b) shows the root-mean-square error, \(\varepsilon_{\mathrm{rms}}\), and the maximum error, \(\varepsilon_{\max}\), evaluated at sites \(1\) and \(16\) for a system size of \(30\) sites subject to $L_{\rm{deph}}$. The decay rates are chosen to be the same as those used in Fig.~\ref{fig:shift} at all sites. As the maximum bond dimension is increased, both error measures remain nearly unchanged, indicating that the results are already converged at a relatively small bond dimension for this particular model.\\

	In Fig.~\ref{fig:nmdep100} and Fig.~\ref{fig:nmdep10000}, we have shown the average $\mu_t$ i.e $\mathbb{E}[\mu_t]$, evolution of the normalized expectation  $\av{X^{[i]}}$ at site $1$ and $16$ in a transverse Ising chain of 30 spins for decay rates, same as those used in Fig.~\ref{fig:shift} at all sites with $\chi_{\max} = 128$, for $N_{\rm traj} = 100$ and $10,000$ respectively. For comparison, we also plot the corresponding expectation values obtained from the numerically exact MPO-based method.\\
	
	\begin{figure}[h]
	    \centering
	    \includegraphics[width=0.95\linewidth]{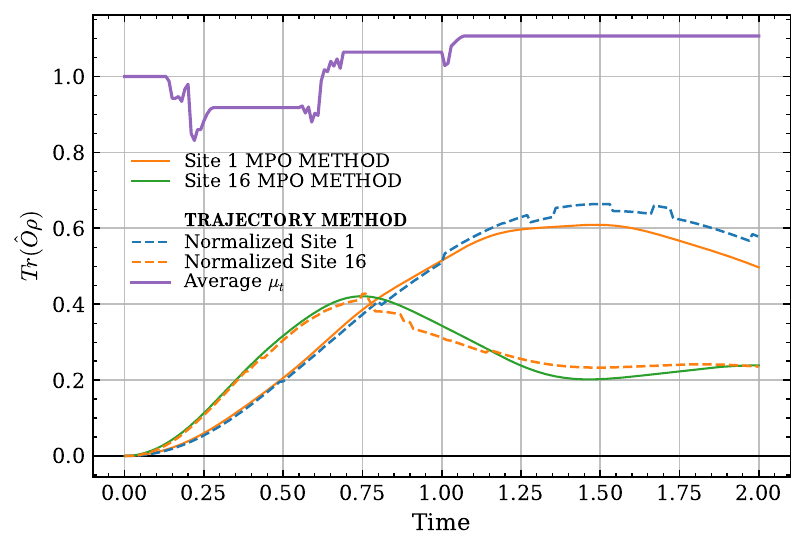}
	    \caption{{Average martingale $\mu_t$ and normalized local expectation values $\langle X^{[i]}\rangle$ for sites $i=1$ and $i=16$ of a 30-spin chain subject to non-Markovian dephasing channel, with decay rates, same as those used in Fig.~\ref{fig:shift} and are applied to all sites. The results are obtained with $\chi_{\max}=128$ and $N_{\rm traj}=100$, while the corresponding numerically exact MPO-based results are shown as a reference.}}
	    \label{fig:nmdep100}
	\end{figure}
	
	\begin{figure}[h]
	\centering
	\includegraphics[width=0.95\linewidth]{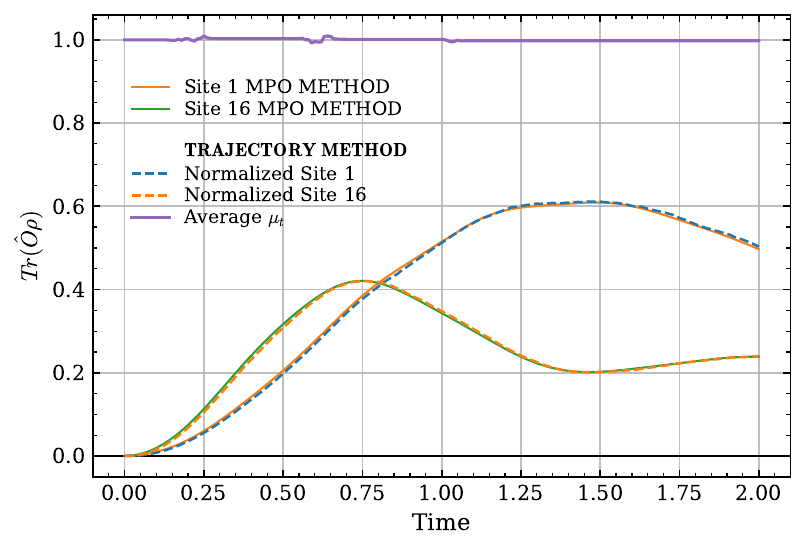}
	\caption{{Average martingale $\mu_t$ and normalized local expectation values
		$\langle X^{[i]}\rangle$ for sites $i=1$ and $i=16$ of a 30-spin chain subject
		to non-Markovian dephasing channel, with decay rates, same as those used in Fig.~\ref{fig:shift} and are applied to all sites. The results are obtained with $\chi_{\max}=128$ and $N_{\rm traj}=10\,000$, while the corresponding numerically exact MPO-based results are shown as a reference.}}
	\label{fig:nmdep10000}
	\end{figure}

	Fig.~\ref{fig:nmexcnmrel} presents average $\mu_t$ and the evolution of the local normalized expectation $\av{X^{[i]}}$ at site $1$ and site $16$ of a 30-spin chain coupled to the environment through the two channels of excitation and relaxation with decay rates, same as those used in Fig.~\ref{fig:shift} at all sites with $\chi_{\max} = 128$ and $N_{\rm traj} = 10,000$. For comparison, corresponding expectation values obtained from the numerically exact MPO-based method are also included.

	\begin{figure}[h]
	\centering
	\includegraphics[width=0.95\linewidth]{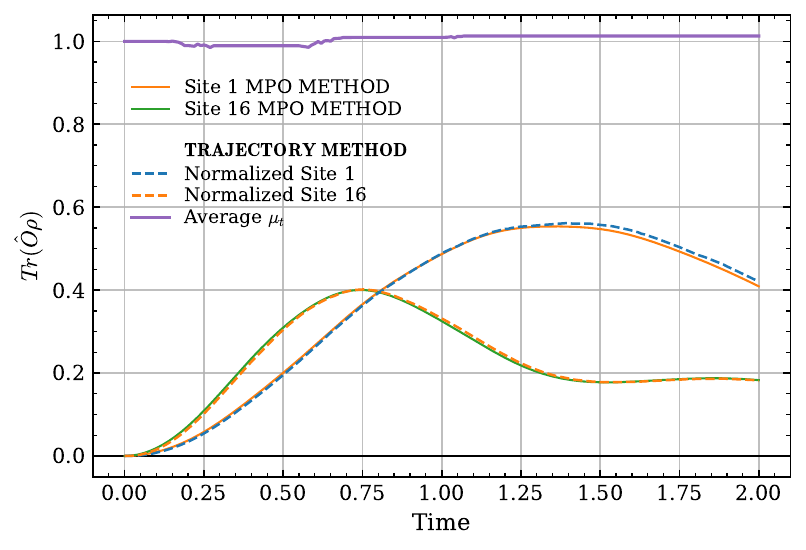}
	\caption{{Average martingale $\mu_t$ and normalized local expectation values
		$\langle X^{[i]}\rangle$ for sites $i=1$ and $i=16$ of a 30-spin chain subject
		to two excitation and relaxation channels, with decay rates, same as those used in Fig.~\ref{fig:shift} and are applied to all sites. The results are obtained with $\chi_{\max}=128$ and $N_{\rm traj}=10\,000$, while the corresponding numerically exact MPO-based results are shown as a reference.}
	}\label{fig:nmexcnmrel}
	\end{figure}

	The trajectory-based results are found to be consistent with the MPO results at the level of the observed dynamical trends, even for $N_{\mathrm{traj}}=100$, suggesting that relatively small trajectory ensembles may already provide useful qualitative information for exploratory studies.\\

	\subsection{Scaling tolerance and `Influence radius':} 
	\label{sec:infrad}
	We investigated the range of decay rates over which the proposed method remains accurate with as few as $N_{\rm traj}=100$ stochastic trajectories. we noticed that major errors of the previous TJM method are either $O(\delta t^2)$ or $O(\delta t^3)$ and ``for 2TDVP the projection error is exactly zero if we consider Hamiltonians with only nearest neighbor interactions" \cite{sander2025largescalestochasticsimulationopen, Haegeman2016, Paeckel2019}. While the error due to the martingale weight scales with number of sites the non markovian noise acts on (${\rm N}_g$) and also the area under shift($C_t$) and can be quantified by,
	{
	\begin{eqnarray}
	{\rm G} &=& \int \sum_{s,l} (r_{s,l}(t) - \gamma_{s,l}(t)) || L_{s,l} \k{\psi(t)}||^2 dt\, \nonumber \\ &=&\, {\rm N}_g\cdot\int C_t\cdot dt,
	\end{eqnarray}
	
as for each site $\sum_l L_l^\dagger L_l = \mathbb{I}$ and $r_s(t) - \gamma_s(t) =C_t\,\, \forall s$},	
where $s$ is site index, $l$ is channel index. For instance, the decay rate used in Fig.~\ref{fig:shift} corresponds to ${\rm G}/{\rm N}_g \simeq 2^{-5.4}$. In Figure ~\ref{fig:influence}(a) we plotted the RMS-error and Max-error with ${\rm G}/{\rm N}_g$ by varying a scale factor $s$ in $\gamma_s(t) = s * \Big[\gamma_\infty - B e^{-f(t)}\sin(\omega t)\Big]$ , while keeping all other parameters identical to those used in Fig.~\ref{fig:shift}. The dephasing Lindblad operator $L_{\rm deph}$  applied on all sites of a 6-site chain(${\rm N}_g = 6$), with $\chi_{\max}=128$ and $N_{\rm traj}=100$. As ${\rm G}$ increases, the variance associated with the martingale weights grows, eventually leading to errors that exceed the acceptable accuracy threshold. Therefore to effectively simulate very large scale many body system, we introduce the concept of \textbf{`Influence Radius'}. \\
	
	\begin{figure*}[t]
		\centering
		\begin{subfigure}{0.48\linewidth}
			\centering
			\includegraphics[width=\linewidth]{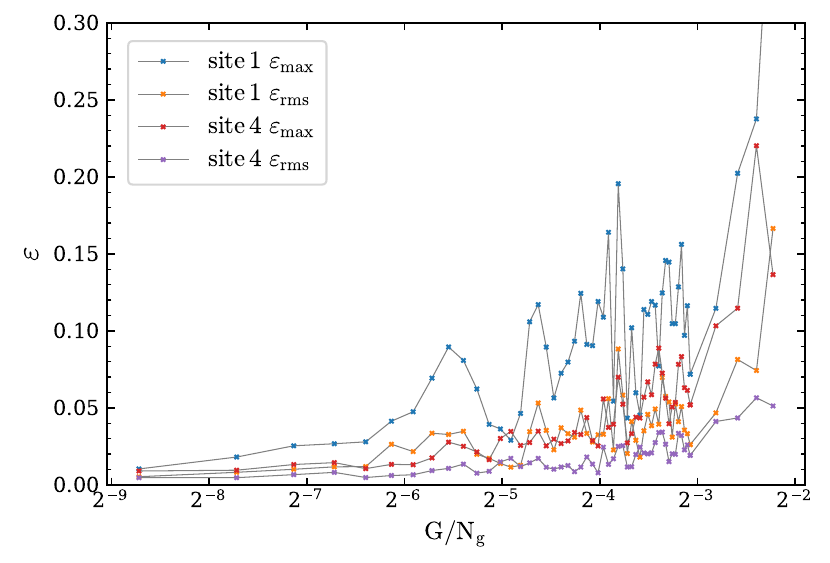}
		\end{subfigure}%
		\hfill
		\begin{subfigure}{0.49\linewidth}
			\centering
			\includegraphics[width=\linewidth]{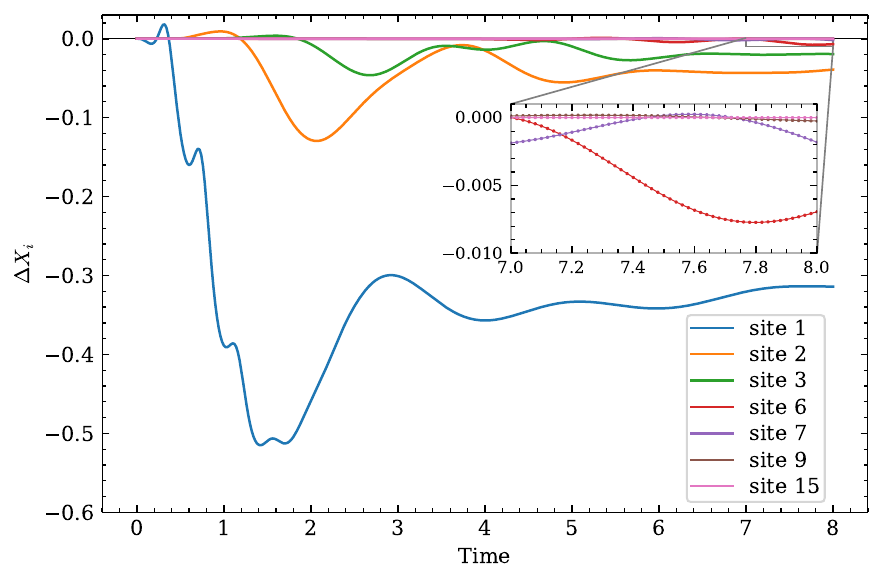}
		\end{subfigure}
		\caption{{(a) {$\varepsilon_{\mathrm{rms}}$} and 
				{$\varepsilon_{\max}$} versus   ${\rm G}/{\rm N}_g$  , obtained by scaling
				$s$ in $\gamma_s(t)$ while keeping all other parameters as used in Fig.~\ref{fig:shift}. $L_{\rm deph}$ is applied on all sites of a 6-site chain, with $\chi_{\max}=128$ and $N_{\rm traj}=100$.
				(b) $\Delta X_i$, {at different sites, for $N = 15$.}, obtained from the MPO-based exact solution under dephasing
				noise. In configuration 1, only site $1$ has the non-Markovian rate
				$\gamma(t)$ of Fig.~\ref{fig:shift}, whereas all other sites have
				$\gamma=0.1$; in configuration 2, all sites have $\gamma=0.1$. The decay of
				$\Delta X_i$ with distance from the locally perturbed site motivates the `influence radius' construction.}}
		\label{fig:influence}
	\end{figure*} 
	\begin{figure*}[t]
		\centering
		\begin{subfigure}{0.48\linewidth}
			\centering
			\includegraphics[width=\linewidth]{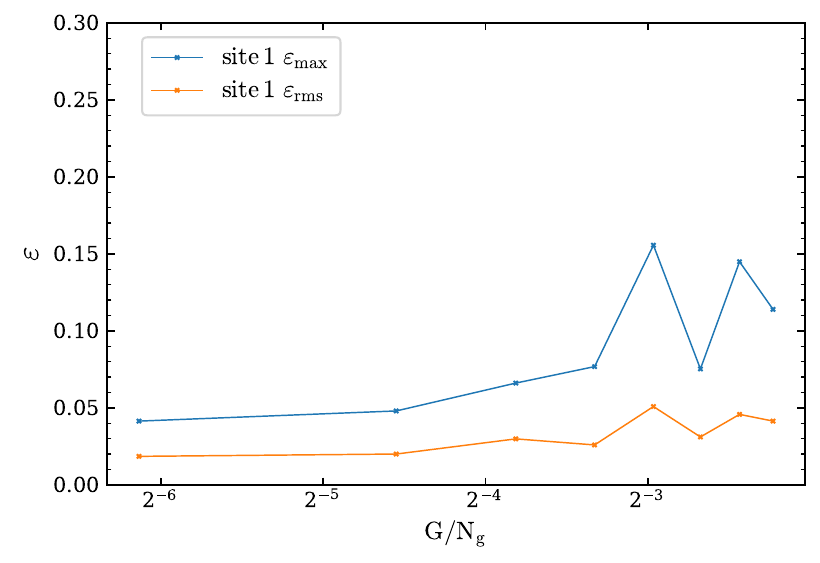}
		\end{subfigure}%
		\hfill
		\begin{subfigure}{0.48\linewidth}
			\centering
			\includegraphics[width=\linewidth]{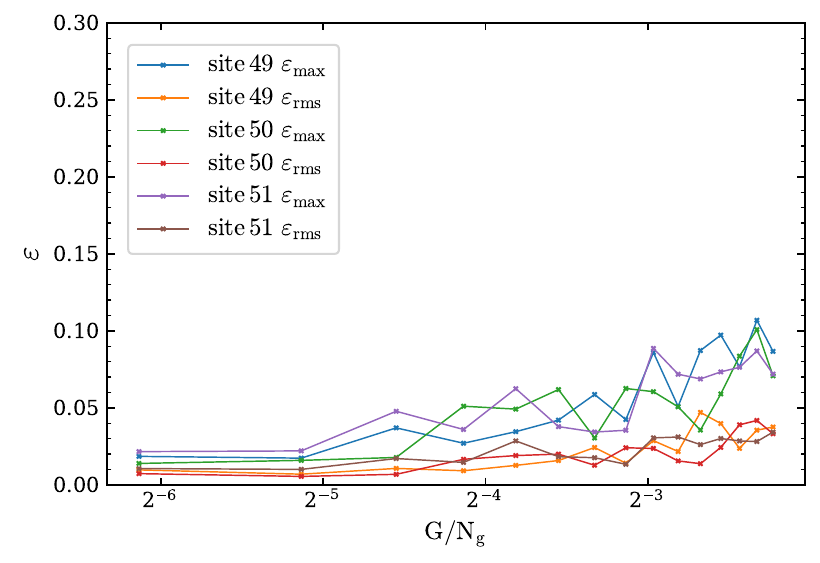}
		\end{subfigure}
		\caption{{$\varepsilon_{\mathrm{rms}}$} and {$\varepsilon_{\max}$} as functions of ${\rm G}/{\rm N}_g$ for a 100-site chain, obtained by varying $s$ in $\gamma_s(t)$, with all other parameters fixed as in Fig.~\ref{fig:shift}. Panels (a) and {(b) correspond to dephasing applied at the boundary site $1$ (${\rm N}_g = 1$)and at the central sites $49$, $50$, and $51$ (${\rm N}_g = 3$), respectively. The simulations use $\chi_{\max}=128$ and $N_{\rm traj}=100$.
		}}
		\label{fig:100site}
	\end{figure*}
	To motivate the concept, Fig.~\ref{fig:influence}(b) compares two different configuration of decay rates subject to dephasing noise ($L_{\rm{deph}}$). First is the non markovian $\gamma = \gamma(t)$ as in Fig.~\ref{fig:shift} at site $1$, while all remaining sites have a constant Markovian decay rate $\gamma=0.1$. In the second configuration consists $\gamma = 0.1$ for all sites.  We use the MPO-based numerically exact solution method to plot the difference between expectation values of two configurations $\Delta X = \av{X^{[i]}}_{\rm config 1} - \av{X^{[i]}}_{\rm config 2}$ at {different sites, for $N = 15$.} \\
 
	We observe that the influence of the non-Markovian noise requires a finite propagation time to reach distant sites and is progressively attenuated with distance. Consequently, for finite-time evolution, the expectation value of a local observable at site $s$ depends appreciably only on the decay rates within a finite spatial neighborhood, extending up to approximately site $s+r$. Beyond this range, the contribution to the observable at site $s$ falls within the prescribed error tolerance. Here, $r$ defines the finite `Influence radius'. For a prescribed accuracy and fixed final time, it is expected to be independent of the total system size.\\
	 
	Thus, based on the numerical results for this particular model, we propose that, for the evaluation of a local observable, the martingale weight need not to be constructed using the decay rates, or equivalently the shifts, over the entire system. Instead, for a prescribed accuracy and fixed final time, it is sufficient to include only those decay rates within the observables' finite `Influence radius'. \\

	\textbf{Application to 100 site: }\\
	To demonstrate the effectiveness of the influence-martingale method with a finite `influence radius' in large systems, we consider two representative configurations of non-Markovian decay rates in a 100-site spin chain subject to dephasing noise\cite{Finsterholzl, Han:2022pvt, Yao2020, Danaci2019gdp}. In the first configuration, referred to as \textit{boundary localized memory}, the time-dependent decay rate $\gamma(t)$(same as in Fig.~\ref{fig:shift}) is applied only at one boundary site, while all other sites are assigned a constant Markovian decay rate $\gamma=0.1$. In the second configuration, referred to as \textit{random localized memory}, the time-dependent decay rate $\gamma(t)$ is applied at one or more randomly selected sites, while the remaining sites again have $\gamma=0.1$.   

	In Fig.~\ref{fig:100site}(a) and \ref{fig:100site}(b), we show the RMS error and maximum error against ${\rm G}/{\rm N}_g$. The data are obtained by varying the scale factor $s$ in $\gamma_s(t)=s[\gamma_\infty-B e^{-f(t)}\sin(\omega t)]$, while keeping all other parameters identical to those used in Fig.~\ref{fig:shift}. In the first configuration, the dephasing Lindblad operator $L_{\rm deph}$ with the time-dependent decay rate $\gamma_s(t)$ is applied at the boundary site $1$ of a 100-site chain(${\rm N}_g = 1$). In the second configuration, the same non-Markovian dephasing noise is applied locally at sites $49$, $50$, and $51$ (${\rm N}_g = 3$). In both cases, the simulations are performed with $\chi_{\max}=128$ and $N_{\rm traj}=100$. Both configurations emulate a finite `influence radius': reducing the effective numbers of non-Markovian sites ${\rm N}_g$. As a result, highers values of $\gamma(t)$ can be accommodated without increasing corresponding error.

}

\section{Conclusion}
\label{sec:con}

{
Classical simulation of open quantum system dynamics remains challenging due to the exponential growth of the Hilbert space, the need to accurately capture dissipation and decoherence, and the added complexity of memory effects in the non-Markovian regime. The Tensor-jump method is a massively scalable algorithm for the simulation of time-homogeneous Markovian open quantum systems by means of
tensor networks. In realistic open-system settings, however, the effective dynamics may acquire an explicit time dependence due to structured reservoirs, finite bath correlation times, external driving, or time-dependent system--environment couplings. Thus it is intriguing to explore both the time-inhomogeneous markovian and non-markovian dynamics.\\

The present approach extends the Tensor Jump Method, which embeds stochastic quantum jumps into MPS tensor networks, to an influence martingale approach. This allows for the unraveling of master equations with time-dependent decay rates, including intervals where decay rates become temporarily negative, through the influence martingale formalism. We discussed and benchmarked in detail, the  computational framework of the algorithm. The convergence analysis shows systematic improvement with the number of stochastic trajectories, while the dependence on the maximum MPS
bond dimension is weak for the parameter regime studied here, suggesting weak entanglement growth for the particular model. Results obtained with 100 trajectories, indicate the domain of applicability for a range of decay rates, suggesting possible exploratory research for Markovian and non-Markovian dynamics without consuming massive computational resources. We also identify the integrated shifted jump intensity as a key quantity governing the growth of martingale-weight fluctuations. Long chain simulations with huge number of sites with negative decay may limit the allowed scale of decay profiles per site, within desired accuracy. \\

To fill this gap, and motivated from the attenuation of the effect of decay rates along distant sites, we introduce the concept of `influence radius', which serve as a quantifier of the local extent to which martingale corrections for local observables are needed to be implemented. This quantifier may be used as a control to optimize computation time and error within  tolerable limit, thereby improving scalability in physically relevant large-chain settings. Two physically relevant noise configurations are briefly studied to demonstrate exploratory research. \\

Taken together, the results show that the influence-martingale trajectory scheme provides an efficient and flexible route to non-Markovian many-body open-system dynamics beyond direct density-matrix evolution. The present framework opens up several directions for detailed future investigations in both Markovian and non-Markovian regimes, encompassing diverse physical scenarios, microscopic bath models, noise configurations, and interacting many-body Hamiltonians.
}
 
\section*{Acknowledgment}
S.M. acknowledges financial support from the Council of Scientific and Industrial Research (CSIR), Government of India, through a Junior Research Fellowship (JRF). S.D. acknowledges financial support from the University Grants Commission (UGC), Government of India, through a Junior Research Fellowship (JRF). {We thank the referees for their valuable comments and constructive suggestions.}

\providecommand{\href}[2]{#2}\begingroup\raggedright\endgroup

\end{document}